\documentclass{jfm}
\usepackage{graphicx,color,multirow,float}
\usepackage{epstopdf, epsfig}
\usepackage{bm,verbatim,amsmath, amssymb}
\newcommand{\Ceps}{\beta}
\newcommand{\Cinf}{\beta_{\infty}}
\newcommand{\RL}{{\rm Re}}
\newcommand{\Rl}{{\rm Re}_{\lambda}}
\newcommand{\vep}{\varepsilon}
\newcommand{\eps}{\varepsilon}
\renewcommand{\vec}[1]{\bm{#1}}
\newcommand{\fvec}[1]{\hat{\bm{#1}}}
\newcommand{\bu}{\vec{u}}
\newcommand{\obu}{\overline{\vec{u}}}
\newcommand{\ou}{\overline{u}}
\newcommand{\os}{\overline{s}}
\newcommand{\vx}{\vec{x}}

\newcommand{\vf}{\vec{f}}

\newcommand{\vk}{\vec{k}}

\newcommand{\beq}{\begin{equation}}
\newcommand{\eeq}{\end{equation}}

\newcommand{\red}[1]{\textcolor{red}{#1}}
\newcommand{\ml}[1]{\textcolor{black}{#1}}

\shorttitle{Nonuniversal bounds on the dissipation rate}
\shortauthor{Moritz Linkmann}

\title{
Effects of helicity on dissipation in homogeneous box turbulence$^*$
}
\author{Moritz Linkmann
  \corresp{\email{moritz.linkmann@physik.uni-marburg.de}},
 }
\affiliation{
Fachbereich Physik, Philipps-Universit\"at Marburg, Renthof 6, 35032 Marburg, Germany
}

\begin{document}

\maketitle

\begin{abstract} 
The dimensionless dissipation coefficient $\Ceps=\vep L/U^3$, 
where $\eps$ is the dissipation rate, $U$ the root-mean-square velocity and $L$ 
the characteristic scale of the largest flow structures,
is an important characteristic of statistically stationary 
homogeneous turbulence. In studies of $\Ceps$, the external force is typically
isotropic and large scale, and its helicity $H_f$ either zero or not measured.
Here, we study the dependence of $\Ceps$ on $H_f$ and find that it  
decreases $\Ceps$ by up to 10$\%$ for both isotropic forces and shear flows.  
The numerical finding is supported by static and dynamical upper bound theory.
Both show a relative reduction similar to the numerical results. 
That is, the qualitative and quantitative 
dependence of $\Ceps$ on the helicity of the force is well captured 
by upper bound theory. 
Consequences for the value of the Kolmogorov constant 
and theoretical aspects of 
turbulence control and modelling are discussed in connection with the 
properties of the external force. In particular, the 
eddy viscosity
in large eddy-simulations of homogeneous turbulence 
should be decreased by at least 10$\%$ in the case of strongly helical forcing.  
\end{abstract}


\section{Introduction}
The Richardson-Kolmogorov cascade picture of fully developed turbulence 
relies on the assumption that the dissipation rate of turbulent
kinetic energy remains finite in the limit of vanishing
viscosity, i.e. on the dissipative anomaly \citep{Frisch95,Eyink03}.
The behaviour of the mean dissipation rate $\vep$ as a function of viscosity 
is mostly studied in nondimensional terms 
through the Reynolds-number dependence of the dimensionless dissipation 
factor $\Ceps = \vep
L/U^3$ \citep{Batchelor53}, 
where $U$ denotes the root-mean-square velocity and
$L$ the size of the largest eddies in the flow. 
The dimensionless dissipation rate is not only of interest in turbulence theory,
as it enters adjustable coefficients in turbulence models such as the 
eddy viscosity in the $k$-$\vep$
model \citep{Tannehill97,Goto09}. 
The Smagorinsky constant in large-eddy simulations (LES) 
also depends on $\Ceps$.    
Since its introduction, the question remains as to whether the infinite-Reynolds-number asymptote of 
$\Ceps$ is a universal quantity, i.e. whether it depends on the forces generating the
turbulence and on the boundary conditions \citep{Goto09,Bos07}.
Since $\Ceps$ is related to the Kolmogorov constant $C_K$ \citep{Lumley92}, 
the question of universality concerning $\Ceps$ extends to 
the Kolmogorov constant. The latter has been an open question 
since the inference by Landau against universality of constants
like $C_K$ \citep{Landau59,Frisch95}.
In particular, it is very difficult to
disprove universality for forces acting at one single characteristic scale, 
as is the case for turbulence generated by a uniform 
grid in e.g. a wind tunnel \citep{Frisch95}.  
The present paper examines the universality of $\Ceps$ and $C_K$ 
with respect to external forces which differ in their 
topological properties, namely their respective helicities, 
while acting at the same single characteristic length scale.

The value of $\Ceps$ has been measured in experiments
\citep{Sreenivasan84,Sreenivasan98,Burattini05} as well as in direct numerical
simulations (DNSs)
\citep{Wang96,
Kaneda03,Gotoh02,Donzis05,Bos07,Goto09,Yeung12,McComb15a,Yeung15,Ishihara16}.
Although the experiments differed in the flow configuration and the DNSs in the
properties of the external forcing and the run time, the results are generally
consistent in terms of $\Ceps \leqslant 1$. However, there is significant spread
between the data points for experimental and numerical results alike.
Similarly, experimentally measured values for the Kolmogorov constant $C_K$
resulted in consistent values $C_K \simeq 1.6$ for different flow
configurations albeit with considerable scatter in the data
\citep{Sreenivasan95}. Furthermore, the highest-resolution DNS of 
homogeneous isotropic turbulence carried out so far 
revealed a difference between the numerically and experimentally measured
values of $C_K$, with $C_K = 1.8 \pm 0.1$ obtained numerically
\citep{Ishihara16}.  
In summary, for both
$\Ceps$ and $C_K$ the difference between the measured values is not large
enough to support non-universality, neither is the statistical error small
enough to disprove it. 

Any question of universality, however,
must be taken in the appropriate context, which is here that of
`equilibrium turbulence' \citep{Batchelor53,Vassilicos15}, where 
the maximal inertial flux $\Pi$ equals $\vep$.
There are many flow configurations where the relation $\Pi = \vep$ is
violated, such as in decaying turbulence and for
unsteady flows \citep{Bos07,Valente12,Valente14,Vassilicos15,Bos17}, where the
variation in the Taylor surrogate $L/U^3$ describes variations of $\Pi$ and not of
$\vep$ \citep{McComb10b,Valente14}. In such cases, the value of $\Ceps$ may 
differ from that for equilibrium turbulence for reasons connected with the 
unsteadiness of the flow.  
Therefore the present paper is only concerned with homogeneous
turbulence maintained in a statistically stationary state by large-scale
external forcing. 

Recent numerical results suggest that $\Ceps$ depends on the number density of
stagnation points in the large-scale flow field, i.e.  on topological details
of the large-scale flow \citep{Goto09}.  A dependence of the inertial flux (and
thus $\vep$) on the topology of the flow field had already been inferred by
\citet{Moffatt85,Moffatt14a} through the effect of kinetic helicity on the
nonlinear structure of the Navier-Stokes equations.  The kinetic helicity is
the $L^2$-inner product $(\vec{u},\vec{\omega})$ of the velocity field
$\vec{u}$ and the vorticity field $\vec{\omega}=\nabla \times \vec{u}$. It is
not only a measure of the alignment between velocity and vorticity and a
conserved quantity under Euler evolution, but also a topological invariant of
the Euler equations related to the linking number of infinitesimal vortex lines
\citep{Moffatt69,Moffatt85}. Since an alignment between $\vec{u}$ and
$\vec{\omega}$ results in a depletion of nonlinearity, regions of high helicity
have been conjectured to be related to low levels of dissipation
\citep{Moffatt14a}.  
\ml{Similar conclusions concerning a depletion of energy transfer in presence 
of strong helicity had already been obtained by \citet{Kraichnan73}
based on interactions of helical Fourier modes.}
Although helicity is an inviscid invariant, it does not
have a coercive effect on the dynamics compared to e.g. the enstrophy in
two-dimensional turbulence, because it is in general not sign definite.
However, once the helicity is made sign definite  through a projection
operation, the energy cascade direction is reversed
\citep{Biferale12,Biferale13a} and the corresponding helically projected
Navier-Stokes equations admit globally regular solutions \citep{Biferale13}. 

Owing to its aforementioned connection to nonlinear Navier-Stokes dynamics and
its relevance to atmospheric physics \citep{Lilly86}, the effect of helicity
has been studied in a variety of turbulent flows, including homogeneous
isotropic turbulence
\citep{Chen03a,Chen03b,Kessar15,Stepanov15a,Gledzer15,Sahoo15a,Alexakis17}, rotating
turbulence \citep{Mininni10a,Mininni10b} and the atmospheric boundary layer
\citep{Deusebio14}.  However, the dependence of $\Ceps$ on the helicity of the
external force has never been investigated analytically or numerically.  The
present work aims to close this gap by providing both analytical estimates and
numerical measurements of $\Ceps$ as a function of the helicity of the forcing.
In view of universality, helicity is also a convenient tool to distinguish
between forcing functions while keeping parameters such as characteristic
length and time scales the same.   

Mathematically rigorous bounds for the dissipation rate have been derived from
the existence of weak solutions of the Navier-Stokes equations for a variety of
wall-bounded flows \citep{Howard72,Busse78,Doering94,Nicodemus98,Kerswell98} as
well as for the case of periodic boundary conditions and sufficiently smooth
forcing functions \citep{Childress01,Foias01,Doering02}. Concerning the
dimensionless dissipation coefficient $\Ceps$, \citet{Doering02} derived the
following bound  
      \beq
      \label{eq:upper_bound}
      \Ceps \leqslant \Cinf + \frac{\ml{\gamma}}{\RL_f} \ ,
      \eeq
where $\Cinf$ and $\ml{\gamma}$ are constants depending on 
the forcing
function \citep{Doering02}, and $\RL_f$ a Reynolds number defined with respect
to the characteristic length scale of the external force.  
The value of the upper
bound has been calculated and compared to experimental and numerical data for
different flow configurations \citep{Doering03,Doering05,Rollin11}.  In all
cases the upper bound is approximately an order of magnitude larger than the measured
value. However, for generalisations of Kolmogorov flow where the effect of
different forcing scales has been studied, the predicted variation of $\Cinf$
is in qualitative agreement with numerically obtained values for $\Ceps$
\citep{Rollin11}. In view of universality, following the arguments by
\citet{Frisch95}, a dependence of $\Ceps$ on the forcing band can indeed be
expected.    

The aim of this paper is to demonstrate that the upper bound theory also
captures the {\em quantitative} dependence of $\Cinf$ as a function of the
helicity
of the force independently of its time dependence, in
the sense that it is able to predict non-universal relative values of $\Cinf$ in
agreement with numerical results.  For this purpose bounds for forces which
differ in their level of helicity and dimensionality are calculated explicitly,
and the upper bound theory is extended to include time-dependent forces.  The
main results of this analysis are: (i) Helical forces lead to lower bounds for
$\Cinf$ compared to non-helical forces.  This supports the rationale of
\citet{Moffatt85,Moffatt14a} that a high level of helicity should inhibit the
energy cascade.  (ii) Dynamic forces lead to larger bounds than static forces,
where the value of the bound depends now also on the characteristic time scale
of the force.  A comparison to DNS data then shows that the relative dependence
of $\Cinf$ on helicity as predicted by the upper bound theory is in good
qualitative and quantitative agreement with numerically measured values of
$\Ceps$, and the results are independent of the dynamical details of the force.
The relative values of $\Cinf$ are related to the relative values of
the Kolmogorov constant $C_K$ in order to predict a qualitative and
quantitative dependence of $C_K$ on the 
helicity
of the forcing. 
Finally, the effect helical forces on the Smagorinsky constant in LES 
is discussed. 

This paper is organised as follows. The necessary mathematical concepts are
introduced in sec.~\ref{sec:notation} alongside the statement of the main
problem and a summary of the derivation of the general upper bound by
\citet{Doering02}.  This method is applied to time-dependent forces in section
\ref{sec:time-dependence}, while the helicity dependence of static forces is
studied in section \ref{sec:geometry-dependence}, including implications
for the Kolmogorov constant and the Smagorinsky constant in LES
in sections \ref{sec:Kolmogorov_const} and \ref{sec:LES}, respectively.  A comparison to DNS data is
carried out in section \ref{sec:DNS}. The main results are summarised and
discussed in section \ref{sec:conclusions}.

\section{Background} \label{sec:notation}
The Navier-Stokes equations are considered on a three-dimensional
domain $\Omega = [0,L]^3$ with periodic boundary conditions 
\begin{align}
\label{eq:momentum}
\partial_t \vec{u}&= - \frac{1}{\rho}\nabla P -(\vec{u}\cdot \nabla)\vec{u}
  + \nu \Delta \vec{u} +  \vec{f}  \ , \\
\label{eq:incompr}
&\nabla \cdot \vec{u} = 0 \ ,  
\end{align}
where $\vec{u}(\cdot,t) \in L^2(\Omega)$ is the velocity field, 
$\nu$ the kinematic viscosity, $P$ the pressure, 
$\vec{f}(\cdot,t) \in L^2(\Omega)$ an 
external mechanical force and $\rho$ the density 
which is set to unity for convenience.
The initial conditions are assumed to be sufficiently
well-behaved to allow weak solutions, i.e 
solutions of the corresponding 
integral equation where all derivatives act on test functions, which are by 
definition infinitely many times differentiable. 
In the following such
weak solutions are considered and any occurrence of a derivative acting on $\vec{u}$ is understood
as shorthand notation for $\vec{u}$ integrated against the derivative of a smooth
test function.

\citet{Leray34} 
established the existence of weak solutions of the Navier-Stokes equation
in three spatial dimensions for square-integrable sufficiently regular 
initial conditions and external forces \citep{Ladyshenskaya69,Constantin88a,Doering95,Foias01}. 
These weak solutions are square integrable and the 
existence result is valid for the three-dimensional torus as well as for the whole space 
$\mathbb{R}^3$ with the appropriate boundary conditions.
Regarding the external force, sufficiently regular usually means that the 
Fourier coefficients of the force are square summable 
(or square integrable, in case of $\mathbb{R}^3$) 
at all times and 
\beq
\sup_{t\geqslant 0} ||(-\Delta)^{-1/2} \vf||_2^2 = L^3 \sup_{t\geqslant 0} 
\sum_{\vk\neq 0} \frac{1}{|\vk|^2}|\fvec{f}(\vk,t)|^2 < \infty \ .
\label{eq:forcebound1}
\eeq
Furthermore, the forces must be solenoidal at all times.


For static forces \citet{Doering02} derived an upper bound on $\vep$ 
from weak solutions 
by decomposing 
the force $\vec{f}$ 
into an amplitude $f_0 \in \mathbb{R}$ and a shape function 
$\vec{\phi} \in L^2([0,1]^3)$, such that
\beq
\label{eq:shape}
\vec{f}(\vx) = f_0 \vec{\phi}(\vx / L_{f}) \ ,
\eeq
where $L_f$ is the characteristic scale at which the force is acting. 
The shape function is further restricted by the requirements 
$||\vec{\phi}||_2 =1$ and 
$||\nabla(-\Delta)^{-M}\vec{\phi}||_\infty < \infty$ 
for some $M \in \mathbb{N}$. Such $M$ can always be found, with the minimum 
requirement for $\vec{\phi} \in L^2([0,1]^3)$ being $M > 1$.
A bound for $\vep$ was then derived from the energy inequality 
\begin{align}
\label{eq:energy_ineq}
\vep(t)  
&  = \nu ||\nabla \vec{u}||_2^2 
   \leqslant (\vec{f},\vec{u}) 
   \leqslant  f_0 ||\vec{\phi}||_2 \ ||\vec{u}||_2  \ ,
\end{align}
by taking the inner product of the Navier-Stokes equations with
$(-\Delta)^{-M}\vec{f}$ and integrating over the volume where several
integrations by parts need to be carried out such that all derivatives act on
the force instead of on the velocity field and the resulting inner products are
bounded from above using the Cauchy-Schwarz and H\"older inequalities. Finally
the long-time average $\langle \cdot \rangle_t$ is taken
\footnote{
The time average can be put on rigorous 
mathematical grounds by considering statistical solutions to the 
Navier-Stokes equations \citep{Foias01}. 
}, resulting in   
\beq
f_0 \leqslant \frac{\|\nabla(-\Delta)^{-M}\vec{\phi}\|_\infty 
                    \langle \|\vec{u}\|_2\rangle_t^2}
                   {L_f \|(-\Delta)^{-M/2}\vec{\phi}\|_2^2 } 
           +  \frac{\nu   \|(-\Delta)^{-M+1}\vec{\phi} \|_2 
              \langle\|\vec{u}\|_2\rangle_t}
                   {L_f^2\|(-\Delta)^{-M/2}\vec{\phi}\|_2^2 } \  .
\eeq
Substitution of the upper bound for $f_0$ into Eq.~\eqref{eq:energy_ineq} and subsequent rearrangement
then yields the following upper bound for $\Ceps$
\beq
\Ceps = \Ceps[\vec{\phi}](\RL_f) \equiv \frac{\vep L_f}{U^3} \leqslant \Cinf + \frac{\ml{\gamma}}{\RL_f} \ , 
\eeq
where $U = \langle ||\vec{u}||^2_2\rangle_t^{1/2}$ and 
\begin{align} \label{eq:cinf_phi}
\Cinf = \Cinf[\vec{\phi}] \equiv \frac{\|\nabla(-\Delta)^{-M}\vec{\phi}\|_\infty \|\vec{\phi}\|_2 }
            {\|(-\Delta)^{-M/2}\vec{\phi}\|_2^2 }  
\quad \text{and} \quad
\ml{\gamma} = \ml{\gamma}[\vec{\phi}] \equiv\frac{ \|(-\Delta)^{-M+1}\vec{\phi} \|_2 \|\vec{\phi}\|_2 }{\|(-\Delta)^{-M/2}\vec{\phi}\|_2^2} \ , 
\end{align}
hence both $\Cinf$ and $\ml{\gamma}$ are functionals of the shape function $\vec{\phi}$.
Here, it is important to observe that unlike $\Cinf$, $\ml{\gamma}$ depends only on
space-averaged quantities and is therefore fully described by the (spatial)
regularity of the shape function, while $\Cinf$ 
\ml{is dominated by its local structure.}
The latter is brought about through $\Cinf$ depending on the
$L^\infty$-norm of the shape function, which involves single-point values.   

\section{Time-dependent forces} \label{sec:time-dependence}

The first task is to extend the results of \citet{Doering02} to time-dependent
forces.  If, as above, the inner product of all terms in the Navier-Stokes
equation with $(-\Delta)^{-M}\vec{f}$ is taken, an extra term arises on the left-hand side
which does not necessarily vanish in the long-time average 
\begin{align}
\label{eq:innerprodtimedep}
- \langle ((-\Delta)^{-M}\partial_t f_i, u_i) \rangle_t
& = \langle (u_i, u_j\partial_j(-\Delta)^{-M}f_i) \rangle_t 
+ \nu \langle ((-\Delta)^{-M}f_i,\Delta u_i) \rangle_t 
\nonumber \\ & \quad 
+ \langle ((-\Delta)^{-M}f_i, f_i) \rangle_t \ .
\end{align}
The main obstacle for an estimation of $\Ceps$ for time-dependent forces thus
lies in that the new term on the left-hand side of Eq.~\eqref{eq:innerprodtimedep} may not
be bounded. This would occur were $\vec{f}$ rough in time. In order to proceed,
$\vec{f}$ could either be assumed to be temporally sufficiently well behaved,
i.e. 
$\vec{f}(\vx, \cdot) \in H^1([0,\infty))$, 
or convoluted with
a filter kernel $G^\tau \in H^\infty([0,\infty))$ such that 
$(G^\tau * \vec{f})(\vx, \cdot) \in H^1([0,\infty))$.  
The latter approach introduces
a time scale $\tau$, which will turn out to be useful in the assessment of the
resulting upper bound of $\Ceps$.  Therefore, instead of using
Eq.~\eqref{eq:innerprodtimedep}, before taking the inner products the force is
smoothed by convolution with $G^\tau$, resulting in 
\begin{align}
\label{eq:innerprodtimedepfilt}
- \langle ((-\Delta)^{-M}\partial_t (G^\tau * f_i), u_i) \rangle_t
& = \langle (u_i, u_j\partial_j(-\Delta)^{-M}G^\tau * f_i) \rangle_t 
+ \nu \langle ((-\Delta)^{-M}G^\tau * f_i,\Delta u_i) \rangle_t 
\nonumber \\ & \quad 
+ \langle ((-\Delta)^{-M}G^\tau * f_i, f_i) \rangle_t \ .
\end{align}
After some intermediate steps involving estimates of $G^\tau$ and 
its time derivative which can be found in Appendix \ref{app:timedep},
one obtains
\begin{align} \label{eq:cinf_phi_timedep}
\Cinf & = \frac{\langle ||\nabla(-\Delta)^{-M}\vec{\phi}||_\infty \rangle_t \langle ||\vec{\phi}||_2 \rangle_t }
            {\langle ||(-\Delta)^{-M/2}\vec{\phi}||_2^2 \rangle_t} 
+ \frac{\omega_f}{\omega}
  \frac{\langle ||(-\Delta)^{-M}\vec{\phi}||_2\rangle_t \langle ||\vec{\phi}||_2 \rangle_t}
       {\langle ||(-\Delta)^{-M/2}\vec{\phi}||_2^2 \rangle_t} \ , 
\end{align}
with $\omega = U/L_f = 1/T$ denoting the frequency corresponding to the
forcing-scale eddy turnover time and $\omega_f \leqslant 1/\tau$ the
characteristic frequency of the smoothed forcing, with $\tau$ being set by the
filter width.  For static forcing $\omega_f = 0$, the time averages in the
definitions of the coefficients $\Cinf$ and $\ml{\gamma}$ can be omitted, and   the forms
of $\Cinf$ and $\ml{\gamma}$ as in Eq.~\eqref{eq:cinf_phi} are recovered.  Dynamic forces
can thus be expected to yield larger bounds due to the extra term in
Eq.~\eqref{eq:cinf_phi} which occurs only for time-dependent forces.  This may
imply that the bound becomes less tight for dynamic forces but it could also
indicate that the value of $\Ceps$ for dynamic forces may be larger than for
static forces. This point will be further assessed in Section \ref{sec:DNS}
using results from numerical simulations. 

\section{Dependence of $\beta$ on the helicity of the force} 
\label{sec:geometry-dependence}
In order to highlight the influence of the helicity  of the force 
on the upper bound of $\Ceps$, the coefficients $\Cinf$ and $\ml{\gamma}$ given in 
Eq.~\eqref{eq:cinf_phi} are calculated explicitly for static forcing 
functions which differ in 
the helicity of their corresponding shape functions. 
For this purpose we consider two shape functions which 
are eigenfunctions of the curl operator
\begin{align}
\label{eq:force_shapepl}
\vec{\phi}^{(1)} = \frac{1}{\sqrt{A^2+B^2+C^2}} 
 \begin{pmatrix}
    A\sin{2\pi z} + C\cos{2\pi y} \\
    B\sin{2\pi x} + A\cos{2\pi z} \\
    C\sin{2\pi y} + B\cos{2\pi x}
  \end{pmatrix} \ .
\end{align}
and 
\begin{align}
\label{eq:force_shapemin}
\vec{\phi}^{(-1)} = \frac{1}{\sqrt{A^2+B^2+C^2}} 
 \begin{pmatrix}
    A\cos{2\pi z} + C\sin{2\pi y} \\
    B\cos{2\pi x} + A\sin{2\pi z} \\
    C\cos{2\pi y} + B\sin{2\pi x}
  \end{pmatrix} \ .
\end{align}
where $A, B, C \in \mathbb{R}$ and $||\vec{\phi}^{(\pm 1)}||_2=1$,
see Appendix \ref{app:norms} for further details. 
These shape functions are by construction fully helical, as their relative
helicity is given by
\beq
\rho_{\vec{\phi}^{(\pm 1)}} = \frac{(\vec{\phi}^{(\pm 1)},\nabla \times \vec{\phi}^{(\pm 1)})}{\|\nabla \times \vec{\phi}^{(\pm 1)}\|\|\vec{\phi}^{(\pm 1)}\|} 
                       = \pm\frac{\|\vec{\phi}^{(\pm 1)}\|_2^2}{\|\vec{\phi}^{(\pm 1)}\|_2^2} =\pm 1 \ , 
\eeq
as $\vec{\phi}^{(\pm 1)}$ are eigenfunctions of the curl operator with 
eigenvalues one and minus one, respectively. The latter also 
implies that $\vec{\phi}^{(1)}$ and $\vec{\phi}^{(-1)}$ are orthogonal
with respect to the $L^2$ inner product. 
A shape function $\vec{\phi}^{(\rho_{\vec{\phi}})}$ of arbitrary relative helicity $\rho_{\vec{\phi}}$ 
is then constructed by suitable linear  
combination of $\vec{\phi}^{(1)}$ and $\vec{\phi}^{(-1)}$
\beq
\label{eq:phi_relhel}
\phi^{(\rho_{\vec{\phi}})} = \sqrt{\frac{1+\rho_{\vec{\phi}}}{2}} \phi^{(1)} + \sqrt{\frac{1-\rho_{\vec{\phi}}}{2}} \phi^{(-1)} \ .
\eeq
Force functions $\vec{f}^{(\rho_f)}$ of a given relative helicity 
$\rho_f \equiv \rho_{\vec{\phi}}$ 
are then constructed according to equation \eqref{eq:shape}.
A further assessment of the effect of dimensionality can be carried 
out by setting one or two of the coefficients $A, B$ or $C$ to zero. 

Before calculating the values of $\Cinf^{(\rho_f)}$  
corresponding to $\vec{f}^{(\rho_f)}$, certain topological and
geometrical properties of the two functions corresponding to the 
cases $\rho_f =1$ and $\rho_f = 0$ are discussed.   
The Navier-Stokes
equations subject to a fully helical force $\vec{f}^{(1)}$ with $f^{(1)}_0=\nu k_f^2$ 
have an exact `laminar'
solution. 
(Here, laminar refers to vanishing nonlinearity, and does not
necessarily imply a layered structure.) 
This solution is
$\vec{f}^{(1)}$ itself, it is known as Arnol'd-Beltrami-Childress (ABC)
flow \citep{Childress70,Dombre86} and has been studied extensively in
connection with dynamo action in magnetohydrodynamics (MHD).  Depending on the
values of $A,B$ and $C$, $\vec{f}^{(1)}$ has up to eight stagnation points
\citep{Dombre86}.  In contrast, a `laminar' flow given by $\vec{f}^{(0)}$
has only the trivial stagnation points $x=y=z=0$ and $x=y=z=\pi$ independently
of the values of $A,B$ and $C$, see Appendix \ref{app:stagnation}.  The two
functions also differ in terms of their symmetry groups, while the symmetry
group of $\vec{f}^{(1)}$ is isomorphic to $\mathbb{Z}_2 \times \mathbb{Z}_2
\times \mathbb{Z}_2$ \citep{Dombre86}, that of $\vec{f}^{(0)}$ is isomorphic
to $\mathbb{Z}_2 \times \mathbb{Z}_2$, see Appendix \ref{app:stagnation}.    

A dependence of the coefficients $\Cinf^{(\rho_f)}$ and $\ml{\gamma}^{(\rho_f)}$ on 
$\rho_f$ is now obtained 
by straightforward analytical evaluation of the
norms on the right-hand side of \eqref{eq:cinf_phi}.  
Since $\vec{\phi}^{(\rho_f)}$ consist of
trigonometric functions they satisfy $(-\Delta)^{-M}\vec{\phi}^{(\rho_f)} =
\vec{\phi}^{(\rho_f)}/(2\pi)^{2M}$, and the $L^2$-norm of their gradients is calculated
directly 
\begin{align}
\|(-\Delta)^{-M/2} \vec{\phi}^{(\rho_f)}\|_2^2 &= ((-\Delta)^{-M/2}\vec{\phi}^{(\rho_f)}, (-\Delta)^{-M/2}\vec{\phi}^{(\rho_f)}) \nonumber \\ 
& = (\vec{\phi}^{(\rho_f)},(-\Delta)^{-M}\vec{\phi}^{(\rho_f)}) = \frac{(\vec{\phi}^{(\rho_f)},\vec{\phi}^{(\rho_f)})}{(2\pi)^{2M}}  
 = \frac{1}{(2\pi)^{2M}} \ .
\end{align}
The evaluation of $\|\nabla(-\Delta)^{-M}\vec{\phi}^{(\rho_f)}\|_\infty = \|\nabla \vec{\phi}^{(\rho_f)}\|_\infty/(2\pi)^{2M}$  
proceeds explicitly by using the definition of the $L^\infty$ norm
\beq
\label{eq:Linfty_gradients}
\hspace{-.15em}
\|\nabla \vec{\phi}^{(\rho_f)}\|_\infty  
= \sup_{\vec{x} \in [0,1]^3}|\nabla \vec{\phi}^{(\rho_f)}| 
 = \sup_{\vec{x} \in [0,1]^3}(\partial_i \phi^{(\rho_f)}_j \partial_i \phi^{(\rho_f)}_j)^{\frac{1}{2}} , 
\hspace{-.12em}
\eeq
where a sum over repeated indices is implied.
Evaluating the last term in Eq.~\eqref{eq:Linfty_gradients} for 
$\vec{\phi}^{(\rho_f)}$ results in 
\beq
\|\nabla \vec{\phi}^{(\rho_f)}\|_\infty = 2 \pi \left ( \sqrt{\frac{1+\rho_f}{2}} + \sqrt{\frac{1-\rho_f}{2}} \right) \ ,
\eeq
see Appendix \ref{app:norms} for further details. 
The values for the norms are now combined according to 
Eq.~\eqref{eq:cinf_phi}, leading to   
\begin{align}
\label{eq:B_NH}
\Cinf^{(\rho_f)} &= \sqrt{2}\pi \left(\sqrt{1-\rho_f}+\sqrt{1+\rho_f} \right) \ , \\ 
\label{eq:C_NH}
\ml{\gamma}^{(\rho_f)} &= (2\pi)^2 \ .  
\end{align}
From Eq.~\eqref{eq:B_NH} one obtains the following 
expression for the helicity dependence of the asymptote
normalised by the zero-helicity value $\Cinf^{(0)}$ 
\beq
\label{eq:ceps-ratio}
\frac{\Cinf^{(\rho_f)}}{\Cinf^{(0)}} 
= \frac{\sqrt{1+\rho_f} + \sqrt{1-\rho_f}}{2} \leqslant 1 \ ,
\eeq
which implies $\Cinf^{(\rho_f)}/\Cinf^{(0)} \in [1/\sqrt{2},1]$. 
That is, a 
helical large-scale force results in a lower estimate for the
non-dimensional total asymptotic energy dissipation rate compared to a 
non-helical force,
provided the forces are acting on the same single length scale.  In contrast,
the approach to the asymptote is independent of $\rho_f$ following Eq.~\eqref{eq:C_NH}.  
Equation \eqref{eq:ceps-ratio} is the first main result of this paper. 

Since $\Cinf$ is also a measure of the inertial flux of the turbulent cascade
for statistically steady turbulence in the infinite-Reynolds-number limit, it
implies that a high level of helicity has a detrimental effect on the energy
cascade. Thus the results obtained by the upper bound theory are qualitatively
in accord with the predictions by Moffatt concerning the effect of helicity on
turbulence dynamics.    The latter prediction, however, was concerned with the
helicity of the flow and not the forcing, which is assessed here.  It is known
that large-scale helicity injection does not lead to highly helical flows, as
mirror symmetry is quickly recovered at successively smaller scales
\citep{Chen03a,Deusebio14,Kessar15}. 
Hence Eq.~\eqref{eq:ceps-ratio} could perhaps best be viewed in terms of a
large-scale control problem: through an adjustment in the helicity of the
forcing it may be possible to regulate the value of the inertial flux across
scales without having to invoke a depletion of nonlinearity in regions of high
helicity at intermediate or small scales.   

\subsection{Variational approach for bidirectional static forces}
The values for the bounds given in Eqs.~\eqref{eq:B_NH} and \eqref{eq:C_NH} do
not depend on the dimensionality of the force because setting either one or two
of the coefficients $A,B$ or $C$ in Eqs.~\eqref{eq:force_shapepl} and
\eqref{eq:force_shapemin} to zero does not alter the results. 
However, for forces
depending on only one spatial coordinate the upper bounds can be improved
through a generalisation of the variational method developed by
\citep{Doering03} for shear flows with unidirectional force, where the
streamwise component of the Navier-Stokes equations is projected on a suitable
multiplier function. The resulting upper bound on $\Ceps$ is then evaluated by
minimisation over the set of multiplier functions \citep{Doering03,Rollin11}. 

This method is not applicable for three-dimensional (3-D) forces, as an average over the direction
of the force is taken. In order to apply it to the present case, set $A=B=0$
such that $\vec{\phi} = (\phi_x(y), 0, \phi_z(y))$ for $y \in [0,1]$, where
$\phi_x$ and $\phi_y$ are periodic functions on $[0,1]$.  Let $\vec{\psi} =
(\psi_x(y),0,\psi_z(y))$ be a function whose second derivative
$\vec{\Psi}=(\partial_y\psi_x,0,\partial_y\psi_z)$ is square integrable (i.e.
$\vec{\psi} \in H^2([0,1])$)  and which satisfies $(\vec{\psi},\vec{\phi})\neq
0$.  Similar to \citet{Doering03}, consider $\vec{\Phi} \equiv
(-\partial_y^{-1} \phi_x, 0, -\partial_y^{-1} \phi_z)$, such that
$(\vec{\Psi},\vec{\Phi}) = (\partial_y\vec{\psi},-\partial_y \vec{\phi}) =
(\vec{\psi},\vec{\phi})$.  Following the procedure outlined in
Sec.~\ref{sec:notation}, {\em i.e.}~taking the inner product of the
Navier-Stokes equation with $\vec{\psi}$ and integrating by parts, one obtains 
\begin{align}
\label{eq:minimax}
\Ceps \leqslant \min_{\psi} \max_{\tilde{\vec{u}}} &
\left(
\frac{
(\tilde{ \vec{u}},(\tilde{ \vec{u}}\cdot \nabla) \vec{\psi})(\tilde{ \vec{u}},\vec{\phi})}{(\vec{\Psi},\vec{\Phi})} \right. 
\left. + \frac{(\tilde{\vec{u}},\partial_y\vec{\Psi})(\tilde{\vec{u}},\vec{\phi})}{\RL_f(\vec{\Psi},\vec{\Phi})} 
\right)
 \ ,
\end{align}
where $\tilde{\vec{u}}=(u_x,u_y,u_z)=\vec{u}/U$. 
The next step consists of a maximisation over all 
divergence-free normalised vector fields $\tilde{\vec{u}}$.
The inner products in the numerators on the right-hand side of Eq.~\eqref{eq:minimax}
are considered separately, beginning with the inertial term
\begin{align}
(\tilde{ \vec{u}},(\tilde{ \vec{u}}\cdot \nabla) \vec{\psi}) &= 
\int_\Omega d\vec{x} \ u_x u_y\partial_y \psi_x(y) + u_z u_y\partial_y \psi_z(y) 
=\int_\Omega d\vec{x} \ u_x u_y\Psi_x(y) + u_z u_y\Psi_z(y) \nonumber \\ 
&=\int_\Omega d\vec{x} \ \tilde{ \vec{u}} \cdot u_y \vec{\Psi}(y) 
\leqslant \|\vec{\Psi}\|_\infty \|u_y \tilde{ \vec{u}}' \|_1 
= \|\vec{\Psi}\|_\infty \int_\Omega d\vec{x} \ |u_y||\sqrt{u_x^2+u_z^2}| \nonumber \\  
&\leqslant \|\vec{\Psi}\|_\infty \int_\Omega d\vec{x}  \ |u_y|(|u_x|+|u_z|) 
\leqslant  \frac{\|\vec{\Psi}\|_\infty}{2} \int_\Omega d\vec{x} \ u_x^2 + u_z^2 + 2u_y^2 \nonumber \\ 
&= \frac{\|\vec{\Psi}\|_\infty}{2} \left(1+\int_\Omega d\vec{x}  \ u_y^2\right) \ ,    
\end{align}
where $\tilde{ \vec{u}}' = (u_x,0,u_z)$, while the 
monotonicity of the square-root was used in 
$\sqrt{u_x^2 +u_z^2} \leqslant \sqrt{(|u_x|+|u_z|)^2}$
and the triangle inequality in $|u_xu_y|\leqslant (u_x^2 +u_y^2)/2$.
For the viscous term, one obtains
\beq
|(\tilde{ \vec{u}},\Delta^{1/2}\vec{\Psi})|=|(\tilde{ \vec{u}}',\Delta^{1/2}\vec{\Psi})|
\leqslant \|\Delta^{1/2}\vec{\Psi}\|_2 \|\tilde{ \vec{u}}'\|_2
= \|\Delta^{1/2}\vec{\Psi}\|_2 \left(\|u_x\|_2^2 + \|u_z\|_2^2 \right)^{1/2} \ ,
\eeq
since $\Psi_y =0$.
The last term to evaluate is
\begin{align}
|(\tilde{ \vec{u}},\Delta^{1/2}\vec{\Phi})| &= |(\tilde{ \vec{u}},\vec{\phi})| 
=\left|\int_\Omega d\vec{x} \ \phi_x u_x + \phi_z u_z \right| 
\leqslant \left|\int_\Omega d\vec{x} \ \phi_x u_x\right| + \left|\int_\Omega d\vec{x} \ \phi_z u_z \right| \nonumber \\
&\leqslant \|\phi_x\|_2 \|u_x\|_2 + \|\phi_z\|_2 \|u_z\|_2  
\leqslant \|u_x\|_2 +  \|u_z\|_2 \ , 
\end{align}
since the normalisation $\|\vec{\phi}\|_2 =1$ implies $\|\phi_x\|_2 \leqslant 1$ and $\|\phi_z\|_2 \leqslant 1$.
Following the procedure of \citet{Doering03}, set
\beq
\xi^2 = \|u_x\|_2^2 + \|u_z\|_2^2 \ ,
\eeq
such that
\begin{align}
&\|u_y\|_2^2 = 1-\|u_x\|_2^2 + \|u_z\|_2^2 = 1-\xi^2 \ , \\
&(\|u_x\|_2 +  \|u_z\|_2)^2 = \xi^2 + 2 \|u_x\|_2 \|u_z\|_2 \leqslant 2\xi^2 \ , 
\end{align}
where the inequality $2|xy|\leqslant x^2 +y^2$ was used again. Now 
Eq.~\eqref{eq:minimax} can be written as
\beq
\Ceps \leqslant
\min_{\vec{\psi}}\frac{1}{(\vec{\Psi},\vec{\Phi})}
  \max_{\xi \in [0,1]}\left(\frac{\xi(2 -\xi^2)}{\sqrt{2}}\|\vec{\Psi}\|_\infty
                          + \frac{\|\Delta^{1/2}\vec{\Psi}\|_2}{\RL_f}\sqrt{2}\xi^2\right) .
\eeq
For $\RL_f \to \infty$
the maximisation over $\xi$ results in $\xi = \sqrt{2/3}$ and $\max_{\xi \in [0,1]}\xi(2 -\xi^2)/\sqrt{2} = 4/\sqrt{27}$,
such that
\beq
\label{eq:minimax2}
\Cinf \leqslant \min_{\vec{\psi}} \max_{\tilde{\vec{u}}}
\frac{
(\tilde{\vec{u}},(\tilde{\vec{u}}\cdot \nabla) \vec{\psi})(\tilde{ \vec{u}},\Delta^{1/2}\vec{\Phi})}{(\vec{\Psi},\vec{\Phi})}
\leqslant \min_{\vec{\psi}}\frac{1}{(\vec{\Psi},\vec{\Phi})}
\frac{4\|\vec{\Psi}\|_\infty}{\sqrt{27}} \ .
\eeq
The remaining minimisation over the multiplier $\vec{\Psi}$ proceeds by 
minor modifications of the method 
devised  by \citet{Doering03,Rollin11}. For this purpose, consider
\beq
(\vec{\Psi},\vec{\Phi}) =  (\vec{\Psi}, (\vec{\Phi}-\vec{C})) \leqslant \|\vec{\Psi}\|_\infty \| (\vec{\Phi}-\vec{C})\|_1  \ ,
\eeq
for any constant vector $\vec{C}=(C_x,C_y,C_z)$, as $\psi_x$ and $\psi_z$ are
periodic functions with zero mean.  The inequality is saturated if
$\vec{\Phi}-\vec{C}$ and $\vec{\Psi}$ are fully aligned, that is if $\vec{\Psi}$ is a
unit vector pointing in the direction of $\vec{\Phi}-\vec{C}$.  The minimum
over $\vec{\Psi}$ in Eq.~\eqref{eq:minimax2} is therefore realised for
\beq
\min_{C_x,C_z} \int_0^1 dy |\vec{\Phi}-\vec{C}| \ ,
\eeq
from which the following conditions for $\vec{C}$ realising the minimum become
\begin{align}
0&=\frac{\partial}{\partial C_x} \int_0^1 dy \  \sqrt{(\Phi_x - C_x)^2 + \Phi_z - C_z)^2}\Big|_{C_{x,z} = C^{\rm min}_{x,z}} \nonumber \\ 
 &= \int_0^1 \frac{\Phi_x-C^{\rm min}_x}{\sqrt{(\Phi_x - C^{\rm min}_x)^2 + (\Phi_z - C^{\rm min}_z)^2}} \ , \\  
0&=\frac{\partial}{\partial C_z} \int_0^1 dy \  \sqrt{(\Phi_x - C_x)^2 + \Phi_z - C_z)^2}\Big|_{C_{x,z} = C^{\rm min}_{x,z}} \nonumber \\ 
 &= \int_0^1 \frac{\Phi_z-C^{\rm min}_z}{\sqrt{(\Phi_x - C^{\rm min}_x)^2 + (\Phi_z - C^{\rm min}_z)^2}} \ .  
\end{align}
For the periodic functions $\vec{\phi}^{X}$ considered here $C^{\rm min}_x = C^{\rm min}_z =0$ satisfies these
conditions.
Hence the final result for the minimax problem in the limit $\RL_f \to \infty$ is
\beq
\label{eq:minimax_final}
\Cinf \leqslant \min_{\vec{\psi}} \max_{\tilde{\vec{u}}}
\frac{
(\tilde{\vec{u}},(\tilde{\vec{u}}\cdot \nabla) \vec{\psi})(\tilde{ \vec{u}},\Delta^{1/2}\vec{\Phi})}{(\vec{\Psi},\vec{\Phi})}
\leqslant\frac{4}{\sqrt{27}}\frac{1}{\int_0^1 dy |\partial_y^{-1}\vec{\phi}|} \ .
\eeq
The final step consists of an evaluation of the integral on the right-hand side of Eq.~\eqref{eq:minimax_final} for
the 
static shape functions $\vec{\phi}^{(\rho_f)}$ 
considered here for $A=B=0$,
resulting in 
\beq
\label{eq:cinf-variational}
\frac{\Cinf^{(\rho_f)}}{\Cinf^{(0)}} = 
\left(\frac{\pi}{2} \int_0^1 dy \ \sqrt{1-\sqrt{1-\rho_f^2}\sin{(4\pi y)}}\right)^{-1} 
\geqslant \frac{2\sqrt{2}}{\pi} \simeq 0.9003 \ ,
\eeq
see Appendix \ref{app:integrals} for the calculation. 
The ratio between the dissipation factors 
is now larger compared to the previous estimate in Eq.~\eqref{eq:ceps-ratio}
because the minimisation procedure replaces the $L^\infty$-norm of $\nabla^{-1}
\vec{\phi}$ with essentially the $L^1$-norm. For the helical shape functions
$\vec{\phi}^{(\pm 1)}$ one thus expects no effect from the minimisation owing to
the fact that unlike for $\nabla \vec{\phi}^{(\rho_f)}$ with $|\rho_f| < 1$, 
the $L^\infty$-norm of $\nabla \vec{\phi}^{(\pm 1)}$ equals the $L^1$-norm.      

\subsection{Implications for the value of the Kolmogorov constant}
\label{sec:Kolmogorov_const}
The dimensionless dissipation coefficient has a direct 
relation to the Kolmogorov constant $C_K$, 
\ml{since the relation $\vep = \beta U^3/L_f$ can be viewed 
as a special case of Kolmogorov scaling formally extended to the 
turbulence production range \citep{Lumley92}. More precisely, if $u_\ell$ is the 
magnitude of the velocity-field fluctuations 
at scale $\ell$ in the inertial range, then  
Kolmogorov-scaling of the energy spectrum implies
$\vep \ell/u_\ell^3 \sim C_K^{-2/3} = const$. Formally extending this 
scaling to the production range, where $\ell = L_f$ and $u_\ell \simeq U$ would 
yield the desired result, which by consequence implies 
$C_K \sim \Ceps^{-2/3}$. However, this is only justified if $L_f$ lies in the 
inertial range, which is not the case at finite Reynolds number. That is, 
the argument can only be applied in the formal limit of infinite Reynolds
number, where the inertial range extends through all $k \neq 0$. This limit
corresponds to replacing $\Ceps$ with $\Cinf$, which yields
\beq
\label{eq:C_K}
C_K \sim \Cinf^{-2/3} \ . 
\eeq
It is important to point out that this argument does not take into account
that $\eps$ can vary locally, 
a point already made by \citet{Lumley92}. Therefore, the scaling 
given in Eq.~\eqref{eq:C_K} can only be viewed as an approximation. 
}
Equation \eqref{eq:C_K} can now be used to obtain the ratio of  
Kolmogorov constants for helical and non-helical forces 
from the ratio $\Cinf^{(\rho_f)}/\Cinf^{(0)}$ 
\beq
\label{eq:C_K_rel}
\frac{C_K^{(\rho_f)}}{C_K^{(0)}} = \left( \frac{\Cinf^{(\rho_f)}}{\Cinf^{(0)}} \right)^{-\frac{2}{3}} \ .
\eeq 
For 3-D static forces, where the minimisation procedure cannot be applied, 
one obtains the following explicit dependence of 
the relative value of the Kolmogorov constant on the helicity of 
the external forcing from Eq.~\eqref{eq:ceps-ratio} and Eq.~\eqref{eq:C_K_rel}
\beq
\frac{C_K^{(\rho_f)}}{C_K^{(0)}} = 
\left( \sqrt{\frac{1+\rho_f}{8}} + \sqrt{\frac{1-\rho_f}{8}} \right)^{-2/3} \leqslant 2^{1/3} \ . 
\eeq 
For shear flows where the minimisation procedure does apply, 
one obtains
\beq
\frac{C_K^{(\rho_f)}}{C_K^{(0)}} = \left(\frac{\pi}{2}\int_0^1 dy \ \sqrt{1-\sqrt{1-\rho_f^2}\sin{(4\pi y)}} \right)^{2/3} \leqslant \frac{\pi^{2/3}}{2} \ .
\eeq
The estimates 
hence result in the following range of values for the two extreme cases 
\beq
1.07 \simeq \frac{\pi^{2/3}}{2} \leqslant \frac{C_K^{(1)}}{C_K^{(0)}} 
  \leqslant 2^{1/3} \simeq 1.26  \ . 
\eeq

\subsection{Implications for the Smagorinsky constant in LES}
\label{sec:LES}
As mentioned in the Introduction, the value of $\beta$ is not only of 
theoretical
interest because of its relation to the parametrisation of the subgrid scales  
in LES, such as for the Smagorinsky model \citep{Smagorinsky63}. 
The aim of LES is to simulate only the motion at large and intermediate scales, while the 
effect of the small scales is modelled. More precisely, let $\obu$ be the velocity field 
$\bu$ convoluted with a filter kernel $G^\Delta$, where $\Delta$ is the characteristic
filter width: $\obu = G^\Delta * \bu$. The evolution of the filtered field is then 
governed by the following equations
\begin{align} 
\label{eq:momentum-les}
\partial_t \obu&= - \frac{1}{\rho}\nabla \overline{P} -(\obu\cdot \nabla)\obu
  + \nu \Delta \obu +  \vec{f} - \nabla \cdot {\bm \tau}^\Delta \ , \\
\label{eq:incompr-les}
&\nabla \cdot \obu = 0 \ ,  
\end{align}
where $\tau_{ij}^\Delta = \overline{u_i u_j} - \ou_i \ou_j$ is the subgrid-scale 
stress tensor and we assume $\Delta < L_f$ such that $\overline{\vec{f}}= \vec{f}$. 
Since $\tau_{ij}^\Delta$ is not closed in term of $\obu$, it must be modelled. 
The Smagorinsky model for $\tau_{ij}^\Delta$ is based on the observation 
that the mean energy flux in 3-D turbulence proceeds from the large scales to the
small scales, it models the deviatoric part of $\tau_{ij}^\Delta$ as 
\beq
\tau_{ij}^\Delta = 2 (c_S\Delta)^2\sqrt{\os_{ij} \os_{ij}} \os_{ij} \ ,
\eeq
where $\os_{ij} = (\partial_i \ou_j+\partial_j \ou_i)$ is the resolved-scale strain tensor and
$c_S$ the Smagorinsky constant, which is an adjustable parameter. 
Since the subgrid-scale energy transfer at scale $\Delta$ is given by 
\beq
\Pi^\Delta = \os_{ij}\tau_{ij}^\Delta \ , 
\eeq
the Smagorinsky model leads to a {\em pointwise} non-negative subgrid-scale energy 
flux
\beq
\Pi^\Delta = \os_{ij} 2 (c_S\Delta)^2\sqrt{\os_{ij} \os_{ij}} \os_{ij} = 2 (c_S\Delta)^2
\sqrt{\os_{ij} \os_{ij}} \os_{ij} \os_{ij} \geq 0 \ .
\eeq
\ml{
The Smagorinsky constant can be related to $\Cinf$ 
using the estimate by \citet{Lilly67} for the value of the Smagorinsky constant 
for statistically steady homogeneous isotropic turbulence, 
$c_S = (3C_K/2)^{-4/3}/\pi$, in combination with Eq.~\eqref{eq:C_K}
\beq
c_S = \frac{\left(\frac{2}{3C_K} \right)^{4/3}}{\pi} \sim \Cinf^{1/2} \ .
\eeq
}
In terms of the dependence of $c_S$ on $\rho_f$, 
the above scaling results in a relative relation 
between $c_S$ and $\Cinf$
\beq
\frac{c_S(\rho_f)}{c_S(0)} = \left(\frac{\Cinf{(\rho_f)}}{\Cinf(0)}\right)^{1/2} \ ,
\eeq
which implies the following dependence of $c_S$ on the relative helicity 
of the forcing
\beq
\frac{c_S(\rho_f)}{c_S(0)} = \left(
 \frac{\sqrt{1+\rho_f} + \sqrt{1-\rho_f}}{2} \right)^{1/2} \ ,
\eeq
for isotropic forcing and
\beq
\frac{c_S(\rho_f)}{c_S(0)} = 
\left(\frac{\pi}{2} \int_0^1 dy \ \sqrt{1-\sqrt{1-\rho_f^2}\sin{(4\pi y)}}\right)^{-1/2} \ ,
\eeq
for shear flows. 
In summary, the values of $c_S$ decrease for increasing $\rho_f$, and 
in case of a strongly helical force the usual value of 
$c_S \simeq 0.17$ \citep{Lilly67} of the Smagorinsky constant should be 
decreased 
according to the corresponding values of $\Cinf$. Since  
the eddy viscosity $\nu_E = 2(c_S\Delta)^2 \sqrt{\os_{ij} \os_{ij}}$ depends
quadratically on $c_S$, it depends linearly on $\Cinf$, 
which results in a decrease of at least 10$\%$ in case of strongly helical forcing. 

\ml{In the context of subgrid-scale modelling, the effect of helicity is usually
included through an extra model term 
\citep{Yokoi93,Li06,Baerenzung08,Inagaki17}, 
leading to an additional diffusion mechanism in 
the model. Here, the modelling of the unresolved inertial 
dynamics as a dissipative loss is the same and only the 
amount of dissipation is changed depending on the helicity of 
the external force.
\citet{Li06} investigated different subgrid-scale models in {\em a-priori} 
and {\em a-posteriori} analyses of isotropic helical turbulence. 
The effect of the newly introduced terms in helical subgrid-scale
models was found to be quite small. Interestingly, 
the dynamic Smagorinsky model, where the model coefficient is 
adjusted in response to the flow, performed best in comparison with DNS data.
An {\em a-posteriori} analysis of the static Smagorinsky model with $c_S$ adjusted as discussed here could be of interest in this context.  
}


\section{Numerical simulations} \label{sec:DNS}

Equations \eqref{eq:momentum}-\eqref{eq:incompr} are solved numerically in a
three-dimensional periodic domain of length $L_{box}=2 \pi$ using a fully
de-aliased pseudospectral code. 
In order to
assess the influence of helicity, dimensionality and time dependence of the
forcing on the value of the dimensionless dissipation coefficient, DNSs were
carried out using different forcing functions, including the static forces
constructed using the shape functions given in Eqs.~\eqref{eq:force_shapemin} 
and \eqref{eq:force_shapepl} according to Eq.~\eqref{eq:phi_relhel}.  
Simulation series carried out using these static
shape functions are identified by the label S, followed 
by the dimensionality of the force and 
the relative helicity level. Here, the label 1D2C refers to one-dimensional
two-component shape functions where e.g. $A=B=0$ while 3D refers to
three-dimensional forces with $A=B=C$.  Since the different implementations of
time-dependent forcing have little effect on the measured value for $\Ceps$
\citep{Bos07}, it is sufficient to consider only one type of time-dependent
forcing for comparison to the static forces. The time-dependent forcing was
given by a Gaussian distributed $\delta(t)$-correlated stochastic process,
which is particularly suited to the present investigation because it gives
optimal control over both kinetic energy and helicity injection rates.
The helicity of the random force is set by expanding the Fourier modes
$\fvec{f}$ of the force field in a basis consisting of eigenfunctions of the
curl operator \citep{Constantin88,Waleffe92}, i.e. into positively and
negatively helical modes, such that the helicity of the force can be adjusted
exactly at each wavevector \citep{Brandenburg01}.  
Simulation series carried
out using dynamic forcing are identified by the labels D1 and D2, 
followed by the helicity level of the force. All simulations of series S and D2
are carried out using $256^3$ collocation points, while simulations of 
series D1 were carried out using $512^3$ collocation points.    
The force always acts the large scales
$L_f=\pi/k_f$, i.e.~at wavenumbers $k_f \leqslant 2.5$ for runs 
of series D1 and at  $k_f =1$ for all other simulations. 
\ml{For case D2, the random force is equivalent to a 
phase-shifted ABC-flow with randomly chosen phases and values of 
$A$, $B$ and $C$.
}
\\

\noindent All runs are carried out with a
fixed time step $dt$ chosen by the Courant-Friedrichs-Lewy criterion, where in
case of white-in-time forcing $dt$ determines the characteristic frequency of
the force by $\omega_f = 2\pi/dt$.  
According to Eq.~\eqref{eq:cinf_phi_timedep},
white-in-time forcing should therefore lead to a maximal weighting of the extra
contribution to $\Cinf$ originating from the time dependence of the forcing
compared to forces with larger correlation times.
Measurements are taken after the simulations have reached a statistically
stationary state, all simulations are evolved for more than $25$ large-eddy
turnover times in stationary state.  It has been pointed out by \citet{Bos07}
that averaging intervals of more than 10 large-eddy turnover times are
necessary in order to obtain accurate values of $\Ceps$.  
The long runtime of the simulations is particularly important for the present
study in order to distinguish the helicity dependence of the measured values of
$\Ceps$ from the statistical error, resulting in a need to compromise between
achievable runtime and resolution. 
A summary of the numerical details including information on the small-scale
resolution and measured values of $\vep$, $U$, $L$ and $\Ceps$ is given in
Tbl.~\ref{tbl:simulations}.  
\ml{
For comparison purposes with results given in the literature for isotropic turbulence, 
where $\beta_L= \vep L/U^3$, with $L$ being the integral scale, is measured 
instead of $\beta = \vep L_f/U^3$, 
values of $\beta_L$ are also provided in the table. 
For the same reason, 
}
$U$ is calculated as $U=\sqrt{2E/3}$, where $E$ is the time-averaged kinetic energy per unit volume.

\begin{table}
 \begin{center}
\begin{tabular}{lccccccccccccccc}
	Run id & N & $\RL$ & $\Rl$ & $\varepsilon$ & $U$ & $L$ & \ml{$\Ceps$} & \ml{$\delta_{\Ceps}$} & $\Ceps_L$ & $\delta_{\Ceps_L}$
	& $\rho$ & $\rho_f$ & $k_{max}\eta$ & $\frac{\nu}{10^3}$ & $t/T$ \\
  \hline
	D1-0 & 512 & 842 & 162 & 0.11 & 0.61 & 0.97 & 1.56 & 0.03 & 0.48 & 0.01 & 0.004 & 0.0 & 1.27 & $0.705$ & 29 \\ 
	D1-1 & 512 & 846 & 168 & 0.08 & 0.57 & 1.05 & 1.35 & 0.05 & 0.45 & 0.02 & 0.15 & 1.0 & 1.39 & $0.705$ & 27 \\ 
  \hline
	D2-0 & 256 & 584 & 151 & 0.09 & 0.68 & 1.55 & 0.88 & 0.06 & 0.433 & 0.03 & 0.001 & 0.0 & 1.37 & $1.8$ & 111 \\ 
	D2-025 & 256 & 532 & 142 & 0.06 & 0.61 & 1.56 & 0.86 & 0.04 & 0.427 & 0.02 & 0.04 & 0.25 & 1.48 & $1.8$ & 100 \\ 
	D2-05 & 256 & 538 & 146 & 0.06 & 0.61 & 1.59 & 0.84 & 0.05 & 0.42 & 0.03 & 0.08 & 0.5 & 1.50 & $1.8$ & 102 \\ 
	D2-075 & 256 & 535 & 146 & 0.05 & 0.59 & 1.64 & 0.77 & 0.04 & 0.40 & 0.02 & 0.14 & 0.75 & 1.57 & $1.8$ & 98 \\ 
	D2-1 & 256 & 616 & 167 & 0.06 & 0.65 & 1.71 & 0.68 & 0.05 & 0.37 & 0.03 & 0.17 & 1.0 & 1.50 & $1.8$ & 106 \\ 
  \hline
	S3D-0  & 256 & 611 & 143 & 0.02 & 0.41 & 1.50 & 0.95 & 0.02& 0.45 & 0.01 & -0.0008 & 0.0 & 1.26 & $1.0$ & 54 \\ 
	S3D-025  & 256 & 600 & 142 & 0.018 & 0.40 & 1.52 & 0.92 & 0.02& 0.447 & 0.01 & 0.07 & 0.25 & 1.30 & $1.0$ & 60 \\ 
	S3D-05  & 256 & 619 & 148 & 0.016 & 0.39 & 1.58 & 0.85 & 0.02 & 0.426 & 0.01 & 0.11 & 0.5 & 1.34 & $1.0$ & 58 \\ 
	S3D-075  & 256 & 629 & 154 & 0.014 & 0.386 & 1.63 & 0.78 & 0.03& 0.40 & 0.01 & 0.14 & 0.75 & 1.39 & $1.0$ & 54 \\ 
	S3D-1  & 256 & 614 & 156 & 0.01 & 0.37 & 1.67 & 0.72 & 0.02& 0.38 & 0.01 & 0.16 & 1.0 & 1.46 & $1.0$ & 65 \\ 
  \hline
	S1D2C-0  & 256 & 645 & 151 & 0.56 & 1.26 & 1.54 & 0.89 & 0.03& 0.43 & 0.01 & 0.01 & 0.0 & 1.24 & $3.0$ & 115  \\ 
	S1D2C-025  & 256 & 584 & 143 & 0.41 & 1.14 & 1.54 & 0.88 & 0.02 & 0.43 & 0.01 & 0.07 & 0.25 & 1.36 & $3.0$ & 168 \\ 
	S1D2C-05  & 256 & 608 & 150 & 0.37 & 1.13 & 1.62 & 0.80 & 0.02 & 0.41 & 0.01 & 0.11 & 0.5 & 1.40 & $3.0$ & 167 \\ 
	S1D2C-075  & 256 & 615 & 155 & 0.31 & 1.10 & 1.68 & 0.73 & 0.02 & 0.39 & 0.01 & 0.15 & 0.75 & 1.49 & $3.0$ & 157 \\ 
	S1D2C-1  & 256 & 630 & 162 & 0.26 & 1.08 & 1.76 & 0.65 & 0.02 & 0.36 & 0.01 & 0.18 & 1.0 & 1.53 & $3.0$ &  162 \\ 
  \end{tabular}
  \end{center}
 \caption{
 Specifications of the numerical simulations.
 $N$ denotes the number of grid points in each Cartesian coordinate,
 $\RL$ the Reynolds number with respect to the rms
 velocity $U$, the integral scale $L$ and the kinematic viscosity $\nu$,
$\Rl$ the Taylor-scale Reynolds number,
$\vep$ the dissipation rate,
$\Ceps=\vep L_f/U^3$ the dimensionless dissipation rate, 
\ml{$\Ceps_L=\vep L/U^3$ the dimensionless dissipation rate with respect to $L$},
$\delta_{\Ceps}$ \ml{and $\delta_{\Ceps_L}$ the respective} standard errors,
$\rho$ the relative kinetic helicity,
 $\rho_f$ the relative helicity of the forcing, $\eta$ the Kolmogorov
 microscale, $k_{max}$ the highest resolved wavenumber,
 $T=L/U$ the large-eddy turnover time, and $t/T$ the
steady-state run time in units of $T$. 
 The values given for $\varepsilon$, $U$, $L$ and $\rho$ are
ensemble averages, with the ensemble consisting of snapshots taken at intervals
of $T$ 
in order to obtain statistically independent
samples.  The identifiers D and S refer to dynamic and static forces,
respectively. The two sets of simulations using static forces differ in the
dimensionality of the force as indicated by the labels 3D and 1D2C.
 }
 \label{tbl:simulations}
 \end{table}

\subsection{Comparison between numerical and analytical results}
A comparison between the values of the rigorous bounds given in
Eqs.~\eqref{eq:B_NH} and \eqref{eq:cinf-variational} and the measured values
given in table \ref{tbl:simulations}  shows that the measured values are
considerably smaller than the corresponding estimates.  The range of values for
the non-helical 3-D forces  $0.43 \leqslant {\Ceps_L}^{(0)} \leqslant 0.49$
obtained from the present DNSs are consistent with existing data from the
literature for 3-D isotropic turbulence
\citep{Wang96,Kaneda03,Gotoh02,Donzis05,Yeung12,McComb15a,Yeung15,Ishihara16},
and the analytically obtained estimates differ by an order of magnitude from
the measured values. Such a discrepancy between the measured value and the
rigorous estimate has also been obtained for a particular type of dynamic
forcing \citep{Doering05}, given by 
\beq
\label{eq:forcing}
\fvec{f}(\vec{k},t) = 
\begin{cases}
(\varepsilon/2 E_f) \hat{\vec{u}}(\vec{k},t) \quad &\text{for} \quad  0 < \lvert\vec{k}\rvert \leqslant k_f ; \\
  \qquad 0    \quad &\text{otherwise} \ ,
\end{cases}
\eeq  
where $\hat{\vec{f}}(\vec{k},t)$ is the Fourier transform of the force and
$E_f$ the total energy contained in the forcing band.  The rigorous bound
derived by \citet{Doering05} resulted in $\Cinf = 4 \pi \sqrt{\frac{3}{5}}
\simeq 9.73$, which could be tightened to $\Cinf = 2\sqrt{2}\pi$ assuming
Kolmogorov scaling for the energy spectrum, {\em i.e.} interestingly to the
same value as $\Cinf^{(0)}$ obtained here for the static 3-D force.  
\\

\begin{figure}
\centering
	\includegraphics[width=0.49\textwidth]{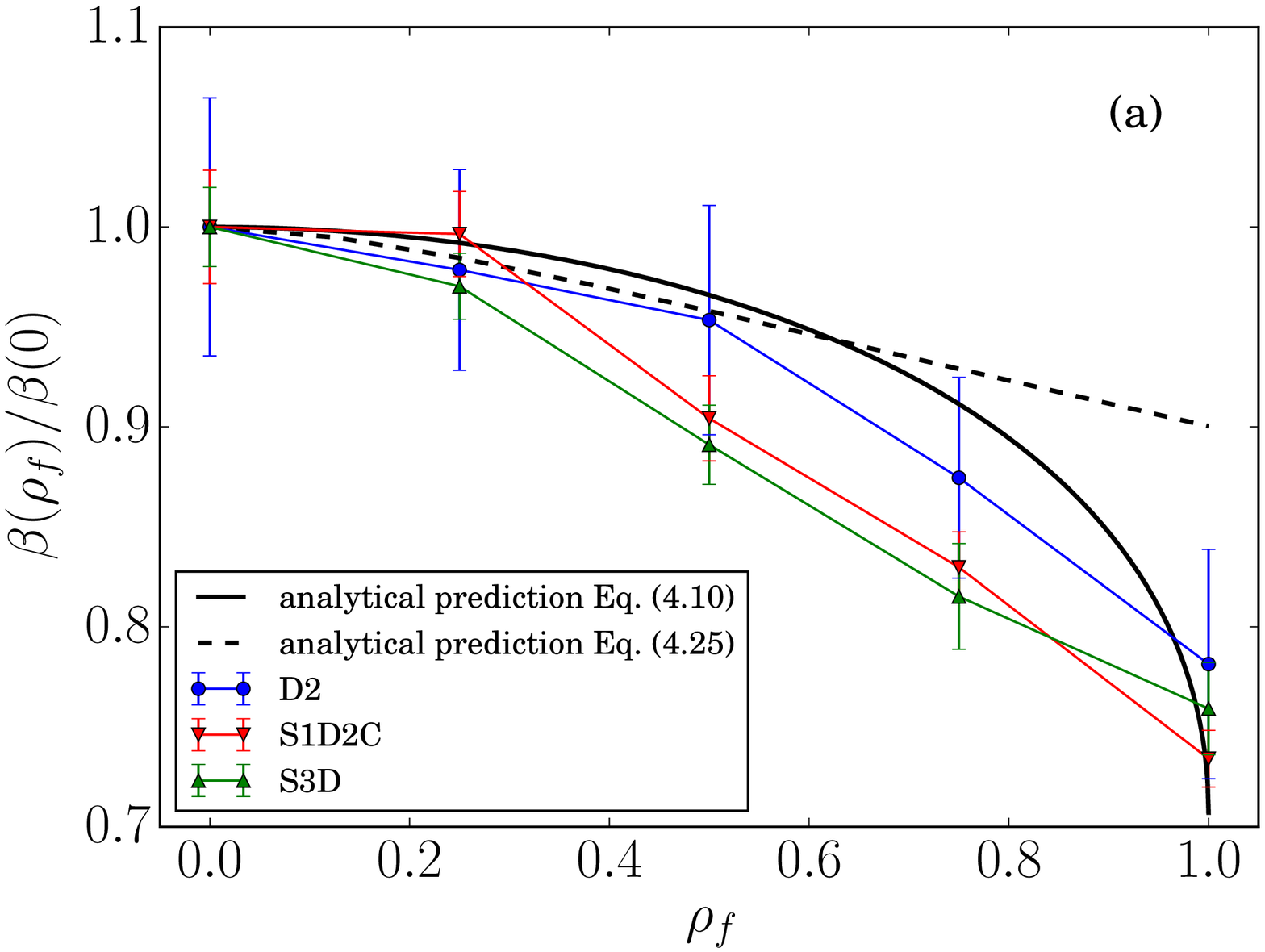}
	\includegraphics[width=0.49\textwidth]{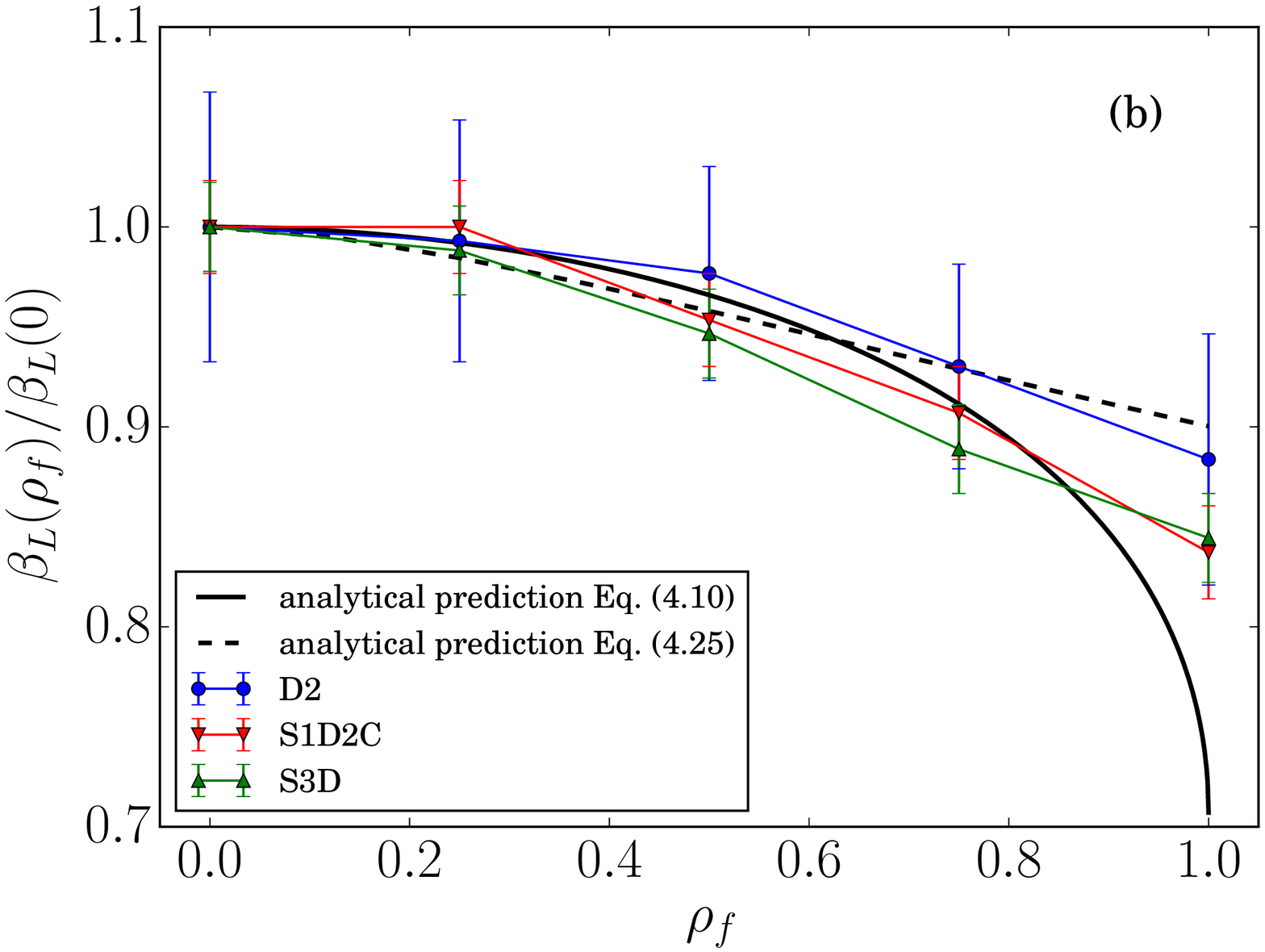}
	\caption{(Colour) Value of $\beta$ as a function of $\rho_f$ normalised by its value at $\rho_f = 0$ 
from datasets D2, S3D and S1D2C compared to the analytical predictions in 
Eqs.~\eqref{eq:ceps-ratio} and \eqref{eq:cinf-variational}.
\ml{(a): $\beta = \vep L_f/U^3$, (b): $\beta_L = \vep L/U^3$, 
where $L$ is the integral scale.
}
}
\label{fig:ceps-vs-relhel}
\end{figure}

Figure \ref{fig:ceps-vs-relhel} presents a comparison of values for
$\Ceps^{(\rho_f)}/\Ceps^{(0)}$ \ml{ (Fig.~\ref{fig:ceps-vs-relhel}(a)) 
and $\Ceps_L^{(\rho_f)}/\Ceps_L^{(0)}$ (Fig.~\ref{fig:ceps-vs-relhel}(b))}
obtained from datasets D2, S3D and S1D2C and the analytical predictions of
Eq.~\eqref{eq:ceps-ratio} and Eq.~\eqref{eq:cinf-variational}.  As can be seen
from the figure, the measured values are in broad agreement between the
different datasets despite the lack of isotropy in case 1D2C and the dynamical
nature of the forcing in case D2.  
\ml{Symmetries specific to the 
choice $A=B=C$ in case S3D have therefore little or no influence on the value of
$\Ceps$.} 
Furthermore, the functional dependence of the ratios
$\Ceps^{(\rho_f)}/\Ceps^{(0)}$ \ml{and $\Ceps_L^{(\rho_f)}/\Ceps_L^{(0)}$} on
$\rho_f$ is consistent with the analytical predictions.  This implies that
although the upper bounds are by an order of magnitude higher than the measured
values, there is a good agreement between the analytical and the numerical
results concerning the ratio $\Ceps^{(\rho_f)}/\Ceps^{(0)}$, which in the limit
$\RL \to \infty$ is predicted to follow Eq.~\eqref{eq:ceps-ratio} for 3-D forces
and Eq.~\eqref{eq:cinf-variational} for shear flows. 
\ml{Differences between the $\rho_f$-dependence of $\Ceps$ and $\Ceps_L$ originate
from a $\rho_f$-dependence of the integral scale, which is discussed  
briefly in Sec.~\ref{sec:integral-scale}.}
\\

The effect of finite Reynolds number on the measurements can be quantified 
through the conventionally band-forced runs D1-0 and D1-1. 
For this purpose, it is
useful to consider the empirical formula obtained by a least-squares fit to a
dataset of $\Ceps_{\ml L}$ resulting from DNSs of stationary homogeneous isotropic
turbulence maintained with the dynamic forcing specified in
Eq.~\eqref{eq:forcing} for $k_f \leqslant 2.5$ \citep{McComb15a}
\beq
\Ceps_{\ml L}^{(0)} = 0.47 + \frac{18.9}{\RL} \ ,
\eeq
which yields $\Ceps_{\ml L}(\RL=834) = 0.49$ in good agreement with the
measured value for run D1-0 shown in table \ref{tbl:simulations}. 
For run D1-1 the same equation is considered after adjusting 
the value of the asymptote according to the aforementioned estimates
\beq
\Ceps_{\ml L}^{(1)} = 0.42 + \frac{18.9}{\RL} \ ,
\eeq
which results in $\Ceps_{\ml L}(\RL=846) = 0.45$, again in good agreement with the
measured value for run D1-1 shown in table \ref{tbl:simulations}. Hence
the ratio 0.9 of the asymptotes and a helicity-independent approach 
to the asymptotes is consistent with the data. 
\\

Concerning a possible influence of the time dependence of the forcing on the
value of $\Cinf$, 
the comparison of values for $\Ceps$ obtained from 
runs D2 and S3D shown in table \ref{tbl:simulations} 
demonstrates that the value of $\Ceps$ is comparable between the dynamically
and the statically forced simulations, provided the forces act at the same
length scales. Furthermore, the ratio $\Ceps^{(\rho_f)}/\Ceps^{(0)}$ appears to
be largely unaffected by the time dependence of the forcing as can be seen 
in Fig.~\ref{fig:ceps-vs-relhel}.  That is, the dynamical
details of the forcing have little influence on the value of $\Ceps$ and
possibly also on that of the asymptote $\Cinf$.  Note that the measured values
of $\Ceps$ for the dynamically forced simulations D1-0 and D1-1 are higher
than those obtained from D2-0 and D2-1, despite the larger Reynolds number
which most probably results from differences in the range of wavenumbers the
force is applied in.  The dependence of $\Ceps$ on the width of the forcing
band was studied analytically and numerically for Kolmogorov flow by
\citet{Rollin11}. The analytical estimates suggested an increase of $\Cinf$
with the width of the forcing band, which was confirmed by DNS results. The
behaviour observed here is consistent with these results, as runs of series 
D2 were forced at $k_f=1$ in order to enable a like-for-like comparison to
the statically forced series S3D runs, while runs of series D1 were forced more
conventionally in the wavenumber band $1\leqslant k_f \leqslant 2.5$ in order
to compare with results in the literature.  \\

In summary, not only the qualitative but more importantly the  relative
quantitative helicity dependence of the measured values of $\Ceps$ is in good
agreement with the helicity dependence of the upper bounds.  Moreover, this
dependence of $\Ceps$ on the helicity 
of the forcing appears to
be independent of its dynamical features.

\subsection{Kolmogorov constant}
Concerning the Kolmogorov constant $C_K$, recent numerical measurements
\citep{Ishihara16} showed that accurate numerical measurements of $C_K$ require
Taylor-scale Reynolds numbers $\Rl \geqslant 700$ and hence very high
resolution DNSs.  Furthermore, numerical results at $\Rl=2297$ requiring
$12288^3$ collocation points revealed a difference between the numerically and
experimentally measured values of $C_K$, with $C_K = 1.8 \pm 0.1$ obtained
numerically \citep{Ishihara16} and $C_K \simeq 1.6$ obtained from experimental
data for several flow configurations \citep{Sreenivasan95}.  The value of the
Kolmogorov constant thus appears still to be an open question, and DNSs at much
higher Reynolds numbers than those carried out in the present paper are
necessary to test any predicted variations for the Kolmogorov constant such as
those presented here. 

\subsection{Further observations}
\label{sec:integral-scale}
As can be seen from table \ref{tbl:simulations}, the integral scale is slightly
larger for helical forces with $L^{(0)}/L^{(1)} \simeq 0.9$ consistently in
all test cases.  
Although a proper interpretation of integral scale is perhaps ambiguous as the
largest scales are dominated by the forcing in the present simulations, the
measurements suggest that helically forced flows consist of larger eddies.
This is expected by the depletion of nonlinearity in regions of high helicity
\citep{Moffatt85,Moffatt14a}.  Although mirror symmetry is generally quickly
recovered at the small scales, \citep{Kraichnan73,Chen03a,Deusebio14,Kessar15}, the high
level of helicity at the large scales diminishes the forward flux of kinetic
energy and hence the efficiency of the kinetic energy cascade leading to less
generation of small-scale turbulent fluctuations \citep{Moffatt14a}.  
\ml{In the decaying case, the same effect results in a delay in the onset of the decay for non-zero helicity \citep{Polifke89}}.
A similar conclusion can be achieved by noting 
that despite comparable large-scale and
Taylor-scale Reynolds numbers, the helically forced turbulent flows are all
better resolved, implying that the Kolmogorov microscale is larger for the
helically forced simulations compared to the non-helically forced runs.     

\begin{figure}
\centering
	\includegraphics[width=\textwidth]{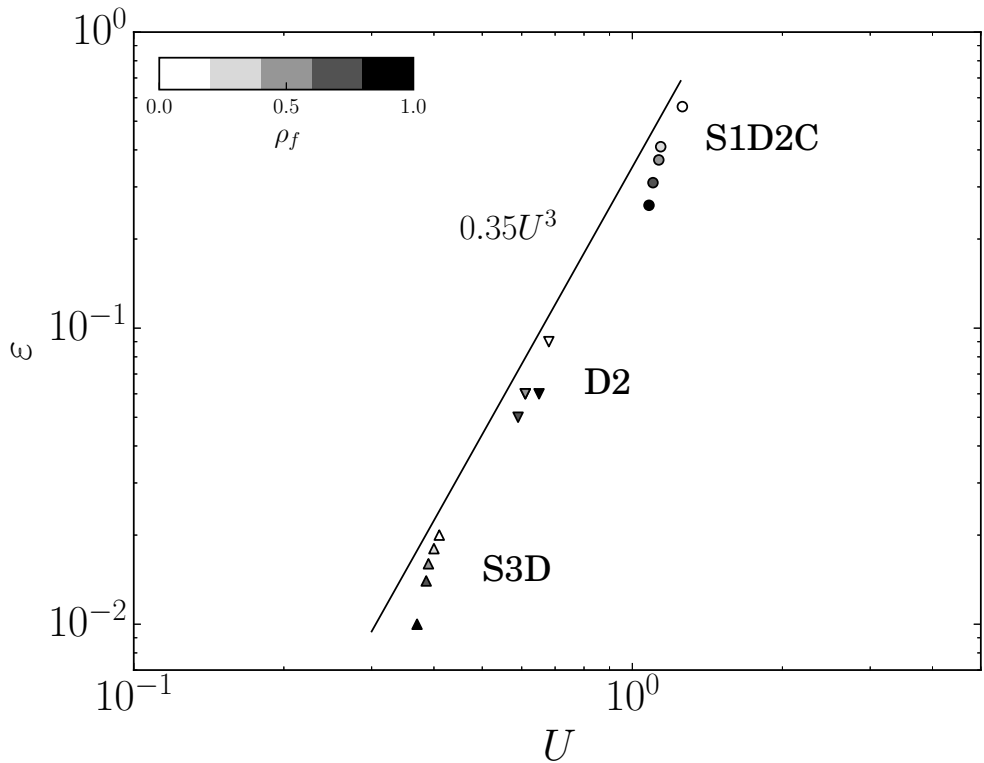}
\caption{
\ml{
Value of $\varepsilon$ as a function of $U$ on a logarithmic scale for datasets S1D2C,
D2 and S3D. The solid line shows the scaling $\varepsilon \sim U^3$,
which corresponds to a fixed value of $\beta$ and $L_f$.  The grey-shading
indicates different values of $\rho_f$. A trend can be observed: For a fixed
value of $U$, the dissipation rate 
decreases with increasing forcing helicity. 
}
}
\label{fig:eps-vs-U}
\end{figure}

\ml{
A reduction in the formation of small-scale structures 
with increasing $\rho_f$ is reminiscent of drag-reducing processes 
in wall-bounded flows.  
More precisely, at a given value of $U$ a decrease in $\varepsilon$ 
in homogeneous turbulence corresponds to a decrease in the wall shear stress 
in wall-bounded flows.  
Such an effect is indeed obtained with increasing $\rho_f$ as shown 
in Fig.~\ref{fig:eps-vs-U}, where $\varepsilon$ is presented as a 
function of $U$. 
It can be quantified through the measure 
\beq
R(\rho_f) = \frac{\beta^{(0)}-\beta^{(\rho_f)}}{\beta^{(0)}} \ ,
\eeq
which is equals the ratio of the corresponding dissipation rates
at fixed $U$. From the analytical and numerical results, one obtains
$R(\rho_f = 1) \simeq 30\%$. 
}

\section{Conclusions}
\label{sec:conclusions}

Upper bounds for the dimensionless dissipation coefficient $\Ceps$ have been
evaluated analytically depending on the relative helicity $\rho_f$ 
of the external forcing. The
main results were: (i) helical forces lead to a lower estimate of the
flux compared to a non-helical force, (ii) a time-dependent force result in a
larger estimate of the flux compared to a static force owing to an extra term 
appearing in the upper bound. The calculated values
of $\Ceps^{(\rho_f)}$ were subsequently compared with values
obtained from DNSs which differed in the helicity level, the time dependence
and the dimensionality of the forcing. The agreement between the theoretically
and numerically obtained values is good {\em concerning the ratio} $\Ceps^{(\rho_f)}/\Ceps^{(0)}$ despite a difference of an order of magnitude between theory
and simulation results concerning the single quantities $\Ceps^{(\rho_f)}$.
Time-dependent forces do not lead to larger values of 
$\Ceps^{(\rho_f)}$ compared to static forces, and the value
of the ratio $\Ceps^{(\rho_f)}/\Ceps^{(0)}$ is comparable between static and
dynamic forces. This indicates that the extra term that appears in the upper
bounds for dynamic forces arises from an analytical difficulty in deriving
tight estimates for dynamic forces and does not carry any relevant information
concerning the value of the energy flux. \\

\noindent         In summary, even though the actual estimates are not very
tight, the upper bound theory captures well the dependence of helicity, i.e. of
a topological property, of the force  on the forward flux of kinetic
energy not only qualitatively but also quantitatively. This result is robust
under differences in the dynamical properties of the forcing.  The forward flux
of energy across the scales can thus be described by the spatial regularity and
the helicity 
of the force, which in principle can be adjusted by
the experimenter. Hence it may be possible to devise a particular type of force
which controls this forward flux of energy, thus leading to a suppression or
enhancement of turbulence and thus of e.g. nonlinear mixing or drag.
The present results also suggest that  detailed knowledge of the
topological properties of a naturally occurring external force field may enable
some predictions about the level of turbulence in a flow.  
Since $\Ceps$ is related to the model coefficient relating the turbulent kinetic
energy to its dissipation rate in the $k$-$\vep$ model and to the eddy viscosity
in LES, the present results may also be useful in practical 
applications concerned with flows subject to helical forces such as in 
atmospheric physics.

\section*{Acknowledgements}
Helpful discussions and suggestions by B.~Eckhardt, C.~Doering, L.~Biferale 
and M.~Buzzicotti are gratefully acknowledged. 
Part of the numerical work
was carried out during a postdoctoral position at the University of Rome `Tor
Vergata' funded through the European Union's Seventh Framework Programme
(FP7/2007-2013) under grant agreement no.~339032.  


\appendix
\section{Time-dependent forcing} \label{app:timedep}
Let $\vec{f} = f_0 g(t) \vec{\phi}(\vec{x}/L_f)$ and consider a Gaussian filter
function $G^\tau$, with characteristic time scale $\tau$.
From the energy inequality one obtains an upper bound for $\vep$ by the same
boundedness argument as in the case of static forcing
\beq
\label{eqapp:epsilon}
\vep \leqslant f_0 \langle g\rangle_t \|\vec{\phi}\|_2 \langle \|\vec{u}\|_2 \rangle_t \ .
\eeq
The next step proceeds similar to the static case by taking the inner product
of all terms in the Navier-Stokes equations 
with $G^\tau * (-\Delta)^{-M}\vec{f}$, and the
arguments concerning the spatial dependence of the force are exactly the same.
Each term in equation \eqref{eq:innerprodtimedepfilt} is now considered separately,
beginning with the new term on the left-hand side
\begin{align}
- \langle ((-\Delta)^{-M}\partial_t (G^\tau * f_i), u_i) \rangle_t 
&= - \langle ((-\Delta)^{-M}\partial_t G^\tau * f_i, u_i) \rangle_t \\
\nonumber 
&= \left \langle \frac{\tau^2}{t^3}((-\Delta)^{-M}G^\tau * f_i, u_i) \right \rangle_t 
\ ,
\end{align}
which results in 
\beq
\langle ((-\Delta)^{-M}\partial_t (G^\tau * f_i), u_i) \rangle_t 
\leqslant f_0 \Big | \left \langle \frac{\tau^2}{t^3} G^\tau * g 
\right \rangle_t \Big | 
\|(-\Delta)^{-M}\vec{\phi}\|_2 \langle \|\vec{u}\|_2 \rangle_t L_f^{2M} \ .
\eeq
For the terms on the right-hand side one obtains
\begin{align}
\langle (u_i, u_j\partial_j(-\Delta)^{-M}G^\tau * f_i) \rangle_t 
& \leqslant f_0  |\langle G^\tau * g \rangle_t| 
  \|\nabla (-\Delta)^{-M}\vec{\phi}\|_\infty (\langle \|\vec{u}\|_2 \rangle_t)^2L_f^{2M-1} \\
\nu \langle ((-\Delta)^{-M}G^\tau * f_i,\Delta u_i) \rangle_t
& \leqslant f_0  |\langle G^\tau * g \rangle_t| 
  \|(-\Delta)^{-M+1}\vec{\phi}\|_2 (\langle \|\vec{u}\|_2 \rangle_t)^2L_f^{2M-2} \\
 \langle ((-\Delta)^{-M}G^\tau * f_i,f_i) \rangle_t 
& = f_0^2  \langle (G^\tau * g)g \rangle_t 
   \|(-\Delta)^{-M/2}\vec{\phi}\|_2^2 L_f^{2M} \ , 
\end{align}
where in the last line $\langle (G^\tau * g)g \rangle_t > 0$.
Hence one obtains the following upper bound 
\begin{align}
f_0 & \leqslant  
U \frac{ |\langle G^\tau * g \rangle_t|}{ \langle (G^\tau * g)g \rangle_t}
\left( 
\frac{\|(-\Delta)^{-M}\vec{\phi}\|_2}{\|(-\Delta)^{-M/2}\vec{\phi}\|_2^2}
+ \frac{U}{L_f} \frac{\|\nabla(-\Delta)^{-M}\vec{\phi}\|_\infty}{\|(-\Delta)^{-M/2}\vec{\phi}\|_2^2}
+ \frac{\nu}{L_f^2} \frac{\|(-\Delta)^{-M+1}\vec{\phi}\|_2}{\|(-\Delta)^{-M/2}\vec{\phi}\|_2^2}
\right) \ ,
\end{align}
which substituted into Eq.~\eqref{eqapp:epsilon} yields 
after some rearrangement a bound on $\Ceps$ 
\begin{align}
\Ceps \leqslant & \
\frac{ |\langle G^\tau * g \rangle_t|\langle g \rangle_t}{ \langle (G^\tau * g)g \rangle_t}
\left(
\frac{|\langle (\tau^2/t^3) G^\tau * g \rangle_t|}{|\langle G^\tau * g \rangle_t|} 
\frac{L_f}{U}
\frac{\|(-\Delta)^{-M}\vec{\phi}\|_2 \|\vec{\phi}\|_2}{\|(-\Delta)^{-M/2}\vec{\phi}\|_2^2}
\right . \nonumber \\
& \qquad \qquad \qquad \qquad 
\left . 
+ \frac{\|\nabla(-\Delta)^{-M}\vec{\phi}\|_\infty \|\vec{\phi}\|_2}{\|(-\Delta)^{-M/2}\vec{\phi}\|_2^2}
+ \frac{1}{\RL} \frac{\|(-\Delta)^{-M+1}\vec{\phi}\|_2 \|\vec{\phi}\|_2}{\|(-\Delta)^{-M/2}\vec{\phi}\|_2^2}
\right) \ .
\end{align}
The summand on the right-hand side of the above inequality can be further approximated by considering 
\beq
\lim_{t \to \infty} (\tau^2/t^3) G^\tau * g = 0 \ , 
\eeq  
since both $G^\tau$ and $g$ are bounded, and
\beq
\lim_{t \to 0} (\tau^2/t^3) G^\tau * g = 0 \ , 
\eeq  
since $G^\tau=\exp{(-\tau^2/t^2)}$ goes to zero faster than any power
for $t \to 0$. 
The average value is thus dominated by the integrand at $t=\tau$ and can be approximated as 
\beq
|\langle (\tau^2/t^3) G^\tau * g \rangle_t| \simeq |\langle G^\tau * g \rangle_t|/\tau \ ,
\eeq
such that with the definitions $\omega_f = 1/\tau$ and $\omega = U/L_f$ one 
obtains Eq.~\eqref{eq:cinf_phi_timedep}.

\section{Evaluation of norms for shape functions $\vec{\phi}^{\rho_f}$.}
\label{app:norms}
The terms to evaluate explicitly are $\|\nabla(-\Delta)^{-M}\vec{\phi}^{\rho_f}\|_\infty$
and $\|\vec{\phi}^{(\rho_f)}\|^2$.
We first establish that the fully helical shape functions are 
normalised to unity 
\begin{align}
\|\vec{\phi}^{(\pm 1)}\|^2_2 &= \frac{1}{|[0,1]^3|} \frac{1}{A^2 + B^2 + C^2}  
 \int_{[0,1]^3} \hspace{-1em} dx \ dy \ dz \ 
\Big(B^2 (\sin{(2\pi x)}^2 + \cos{(2\pi x)}^2) \nonumber \\
& \qquad 
+ C^2 (\sin{(2\pi y)}^2 + \cos{(2\pi y)}^2) 
+ A^2 (\sin{(2\pi z)}^2 + \cos{(2\pi z)}^2) \Big)
= 1 \ .
\end{align}
Since $\vec{\phi}^{(\pm 1)}$ are eigenfunctions of the curl operator, 
they are also orthogonal with respect to the $L^2$-inner product, i.e. $(\vec{\phi}^{(1)}, \vec{\phi}^{(-1)}) =0 $. 
For a shape function with fractional relative helicity we therefore 
obtain
\begin{align}
\|\phi^{(\rho_f)}\|_2^2 &= \left(\sqrt{\frac{1+\rho_f}{2}} \vec{\phi}^{(1)} 
                                     + \sqrt{\frac{1-\rho_f}{2}} \vec{\phi}^{(-1)},
                                       \sqrt{\frac{1+\rho_f}{2}} \vec{\phi}^{(1)} 
                                     + \sqrt{\frac{1-\rho_f}{2}} \vec{\phi}^{(-1)}\right) \nonumber \\ 
                                   &= \frac{1+\rho_f}{2}\|\vec{\phi}^{(1)}\|^2_2 + \frac{1-\rho_f}{2}\|\vec{\phi}^{(-1)}\|^2_2
                                    = 1 \ .
\end{align}
The term $\|\nabla(-\Delta)^{-M}\vec{\phi}^{(\rho_f)}\|_\infty = 
\|\nabla \vec{\phi}^{(\rho_f)}\|_\infty/(2\pi)^{2M}$ is calculated by 
first considering the gradients of the shape functions
\begin{align}
\nabla \vec{\phi}^{(1)} &= \frac{2\pi}{\sqrt{A^2+B^2+C^2}} 
   \begin{pmatrix} 
      0 & B \cos{2\pi x} & -B \sin{2\pi x} \\
      -C \sin{2\pi y} & 0 & C \cos{2\pi y} \\
      A \cos{2\pi z} & -A \sin{2\pi z} & 0 
   \end{pmatrix} \ , \\
\nabla \vec{\phi}^{(-1)} &= \frac{2\pi}{\sqrt{A^2+B^2+C^2}} 
   \begin{pmatrix} 
      0 & -B \sin{2\pi x} & B \cos{2\pi x} \\
      C \cos{2\pi y} & 0 & -C \sin{2\pi y} \\
      -A \sin{2\pi z} & A \cos{2\pi z} & 0 
   \end{pmatrix} \ .
\end{align}
Now the $L^\infty$-norm of $\nabla \vec{\phi}^{(\rho_f)}$ can be calculated. For this purpose, set $\alpha \equiv \sqrt{(1+\rho_f)/2}$ 
and $\gamma \equiv \sqrt{(1-\rho_f)/2}$, such that
\begin{align}
\| & \nabla \vec{\phi}^{(\rho_f)}\|_\infty 
 = \sup_{\vec{x} \in [0,1]^3} \left(\sum_{i,j=1}^3\partial_i \phi^{(\rho_f)}_j \partial_i \phi^{(\rho_f)}_j\right)^{\frac{1}{2}} 
 = \sup_{\vec{x} \in [0,1]^3} \left(\sum_{i,j=1}^3\left[\partial_i \left(\alpha \phi^{(1)}_j +\gamma \phi^{(-1)}_j\right)\right]^2 \right)^{\frac{1}{2}} \nonumber \\
&= \frac{2\pi}{\sqrt{A^2+B^2+C^2}}\sup_{\vec{x} \in [0,1]^3} \left( B^2\left[\left(\alpha \cos{2\pi x}-\gamma \sin{2\pi x}\right)^2 \right. 
                                                                           + \left(\gamma \cos{2\pi x}-\alpha \sin{2\pi x}\right)^2 \right]  \nonumber \\ 
                                      & \left. \qquad  + C^2\left[ \left(\alpha \cos{2\pi y}-\gamma \sin{2\pi y}\right)^2 
                                                                        + \left(\gamma \cos{2\pi y}-\alpha \sin{2\pi y}\right)^2\right] \right. \nonumber \\
                                      & \left. \qquad  + A^2\left[ \left(\alpha \cos{2\pi z}-\gamma \sin{2\pi z}\right)^2 
                                                                        + \left(\gamma \cos{2\pi z}-\alpha \sin{2\pi z}\right)^2\right] \right)^{1/2} \nonumber \\ 
& =  \frac{2\pi}{\sqrt{A^2+B^2+C^2}}\sup_{\vec{x} \in [0,1]^3} \left(B^2 \left[ \alpha^2 + \gamma^2  - 4\alpha \gamma \cos{2\pi x}\sin{2\pi x} \right] \right. \nonumber \\
                                      & \left. \  + C^2 \left[ \alpha^2 + \gamma^2  - 4\alpha \gamma \cos{2\pi y}\sin{2\pi y} \right] 
                                       + A^2 \left[ \alpha^2 + \gamma^2  - 4\alpha \gamma \cos{2\pi z}\sin{2\pi z} \right] \right)^{1/2} . 
\end{align}
Since $\sqrt{a}$ is a monotonic function for $a\in \mathbb{R}$, the supremum is realised at a point 
$\vec{x} = (x,y,z) \in [0,1]^3$ where each summand is maximal. This is the case for $x = y = z = 1/8$ 
since $\cos{\pi/4} = 1/\sqrt{2}$ and $\sin{\pi/4} = -1/\sqrt{2}$, such that 
\begin{align} 
\| \nabla \vec{\phi}^{(\rho_f)}\|_\infty & = 
  \frac{2\pi}{\sqrt{A^2+B^2+C^2}}\left((A^2 + B^2 + C^2) \left[ \alpha^2 + \gamma^2  + 2\alpha \gamma \right] \right)^{1/2} = 2\pi|\alpha + \gamma| \nonumber \\
& = \sqrt{2}\pi \left(\sqrt{1-\rho_f}+\sqrt{1+\rho_f} \right) \ .
\end{align}
Finally, one obtains 
\begin{align}
\|\nabla(-\Delta)^{-M}\vec{\phi}^{(\rho_f)}\|_\infty &=  
  \|\nabla \vec{\phi}^{(\rho_f)}\|_\infty/(2\pi)^{2M} 
= \frac{\sqrt{2}\pi \left(\sqrt{1-\rho_f}+\sqrt{1+\rho_f} \right)}{(2\pi)^{2M}} \ .
\end{align}

\section{Evaluation of the integrals in Eq.~\eqref{eq:minimax_final} for bidirectional static forces}
\label{app:integrals}
Consider the two static forces $\vec{\phi}^{(\pm 1)}$ 
for $A=B=0$. For simplicity we set $C=1$, such that 
\beq
-\partial_y^{-1}\vec{\phi}^{(1)}=\frac{1}{2\pi}
\begin{pmatrix}
-\sin{2\pi y} \\ 0 \\ \cos{2\pi y}
\end{pmatrix}
\qquad \text{ and } \qquad
-\partial_y^{-1}\vec{\phi}^{(-1)}=\frac{1}{2\pi}
\begin{pmatrix}
\cos{2\pi y} \\ 0 \\ -\sin{2\pi y}
\end{pmatrix} \ ,
\eeq
such that 
\beq
-\partial_y^{-1}\vec{\phi}^{(\rho_f)} = \sqrt{\frac{1+\rho_f}{2}} 
   \begin{pmatrix}
    -\sin{2\pi y} \\ 0 \\ \cos{2\pi y}
   \end{pmatrix}
   + \sqrt{\frac{1-\rho_f}{2}}
   \begin{pmatrix}
     \cos{2\pi y} \\ 0 \\ -\sin{2\pi y}
   \end{pmatrix} \ .
\eeq
The evaluation of the integral on the right-hand side of 
Eq.~\eqref{eq:minimax_final} proceeds by explicit calculation. For convenience, set 
$\alpha \equiv \sqrt{(1+\rho_f)/2}$ and $\gamma \equiv \sqrt{(1-\rho_f)/2}$,
such that  
\begin{align}
\int_0^1 \hspace{-0.4em} dy \ |\partial_y^{-1}\vec{\phi}^{(\rho_f)}| & 
=  \frac{1}{2\pi}\int_0^1 \hspace{-0.4em} dy \ \sqrt{(\alpha \cos{2\pi y} - \gamma \sin{2\pi y} )^2 
                                   + (\gamma \cos{2\pi y} - \alpha \sin{2\pi y} )^2 } \nonumber \\
& =  \frac{1}{2\pi}\int_0^1 \hspace{-0.4em} dy \ \sqrt{\alpha^2 + \gamma^2 - 2\alpha\gamma \sin{4\pi y} } \nonumber \\
& = \frac{1}{2\pi}\int_0^1 \hspace{-0.4em} dy \ \sqrt{1-\sqrt{1-\rho_f^2} \sin{4\pi y}} \ , 
\end{align}
where the integrand has no closed-form antiderivative.
For the extreme cases $\rho_f = \pm 1$ and $\rho_f = 0$, one obtains 
\begin{align}
\hspace{-1em} \int_0^1 \hspace{-0.4em} dy \ |\partial_y^{-1}\vec{\phi}^{(\pm 1)}| 
& = \frac{1}{2\pi}\int_0^1 \hspace{-0.4em} dy \ \sqrt{(\sin{2\pi y})^2 + (\cos{2\pi y})^2} = \frac{1}{2 \pi} \ , \\
\hspace{-1em} \int_0^1 \hspace{-0.4em} dy \  |\partial_y^{-1}\vec{\phi}^{(0)}| 
& = \frac{1}{2\pi}\int_0^1 \hspace{-0.4em} dy \ \sqrt{1-\sin{4\pi y}} = \frac{1}{2\pi}\int_0^1 \hspace{-0.4em} dy \ \sqrt{(\cos{2\pi y}-\sin{2\pi y})^2} \nonumber \\
& = \frac{1}{2\pi}\int_0^1 \hspace{-0.4em} dy \ \sqrt{2}|\sin{(2\pi y +\pi/4})| = \frac{1}{2\pi}\int_0^1 \hspace{-0.4em} dy \ \sqrt{2}|\sin{2\pi y}| = \frac{\sqrt{2}}{\pi^2}  .
\end{align}

\section{Stagnation points and symmetries}
\label{app:stagnation}
In this appendix we consider the stagnation points and symmetries of 
a flow corresponding to $\vec{\phi}^{(0)}$, i.e. given by  
\begin{align}
\vec{v}^{(0)} \equiv
\begin{pmatrix}
\dot{x}(t) \\ 
\dot{y}(t) \\ 
\dot{z}(t)  
  \end{pmatrix} = 
\begin{pmatrix}
    A\sin{2\pi z(t)} + C\sin{2\pi y(t)} \\
    B\sin{2\pi x(t)} + A\sin{2\pi z(t)} \\
    C\sin{2\pi y(t)} + B\sin{2\pi x(t)}
  \end{pmatrix} \ , 
\end{align}
on the periodic domain $[0,1)^3$. 
The stagnation points of $\vec{v}^{(0)}$ require $\vec{v}^{(0)} = 0$, however 
\begin{align}
\dot{x}(t) = 0 & \implies A\sin{2\pi z(t)} = - C\sin{2\pi y(t)} \ , \\
\dot{z}(t) = 0 & \implies B\sin{2\pi x(t)} = - C\sin{2\pi y(t)} \ , 
\end{align} 
result in $\dot{y}(t)= -2 C\sin{2\pi y(t)}$. Hence $\vec{v}^{(0)} = 0$ if and only if
$x=y=z=0$ or $x=y=z=\pi$. 
The symmetry group of $\vec{v}^{(0)}$ consists of the following four elements
$\{{\rm id}, \sigma_1, \sigma_2, \sigma_3\}$ where ${\rm id}$ denotes the identity transformation and 
\begin{align}
&\sigma_1(x) = -x, \, \sigma_1(y) = -y, \, \sigma_1(z) = -z, \, \sigma_1(t) = t; \\ 
&\sigma_2(x) = x+\pi, \, \sigma_2(y) = y+\pi, \, \sigma_2(z) = z+\pi, \, \sigma_2(t) = t; \\ 
&\sigma_3(x) = -x-\pi, \, \sigma_3(y) = -y-\pi, \, \sigma_3(z) = -z-\pi, \, \sigma_3(t) = -t.  
\end{align}
Since $\sigma_3=\sigma_1 \circ \sigma_2$, the set $\{{\rm id}, \sigma_1, \sigma_2, \sigma_3\}$ indeed
forms a group. It is isomorphic to the direct product of the cyclic group of two elements $\mathbb{Z}_2$
with itself because $\sigma_i^2 = {\rm id}$ for $i \in \{1,2,3\}$.
  
\bibliographystyle{jfm}
\bibliography{jfm-refs,refs,refs1,refs2,wdm,apj_refs}

\end{document}